\documentclass{article}
\usepackage[utf8]{inputenc}
\usepackage{todonotes}
\usepackage{amsmath,bm}
\usepackage{graphicx}
\usepackage{multirow,array} 
\usepackage{subfig}
\usepackage{colortbl}
\usepackage[round]{natbib}
\usepackage{float}
\usepackage{soul}
\graphicspath{{figures/}}
\usepackage{algorithm2e}
\usepackage{textcomp,graphicx,amssymb}
\usepackage{amsthm}
\usepackage{url}
\usepackage{authblk}
\newcommand{\sump}{\sideset{}{'}\sum}
\newtheorem*{remark}{Remark}

\title{A neural network-based framework for financial model calibration}

\author[1]{\small Shuaiqiang Liu*}
\author[2]{\small Anastasia Borovykh} 
\author[1]{\small Lech A. Grzelak} 
\author[1, 2]{\small Cornelis W. Oosterlee}

\affil[*]{\footnotesize Email: s.liu-4@tudelft.nl}
\affil[1]{\footnotesize Applied Mathematics (DIAM), Delft University of Technology,  Delft, the Netherlands}
\affil[2]{\footnotesize Centrum Wiskunde \& Informatica, Amsterdam, the Netherlands}

\begin{document}

\maketitle

\begin{abstract}
A data-driven approach called CaNN (Calibration Neural Network) is proposed to calibrate financial asset price models using an Artificial Neural Network (ANN).
Determining optimal values of the model parameters is formulated as training hidden neurons within a machine learning framework,
 based on available financial option prices.  The framework consists of two parts: a forward pass in which we train the weights of the ANN off-line, valuing options under many different asset model parameter settings; and a backward pass, in which we evaluate the trained ANN-solver on-line, aiming to find the weights of the neurons in the input layer.
 The rapid on-line learning of implied volatility by ANNs, in combination with the use of an adapted parallel global optimization method, tackles the computation bottleneck and provides a fast and reliable technique for calibrating model parameters while avoiding, as much as possible, getting stuck in local minima. Numerical experiments confirm that this machine-learning framework can be employed to calibrate parameters of high-dimensional stochastic volatility models efficiently and accurately.
 \end{abstract}

\section{Introduction}
Model calibration can be formulated as an inverse problem, where, based on observed output results, the input parameters need to be inferred. 
 Previous work on solving inverse problems includes research on adjoint optimization methods \citep{DENG200816,inverseproblemofoptionpricing_1997}), Bayesian methods \citep{bayesiancalibration2001,bayOption}, and sparsity regularization \citep{sparsity2004}. 

In a financial context, e.g., in the pricing and risk management of financial derivative contracts, asset model calibration  aims at recovering the model parameters of the underlying stochastic differential equations (SDEs) from observed market data. 
In other words, in the case of stocks and financial options, the calibration aims to determine the stock model parameters such that heavily traded, liquid option prices can be 
recovered by the mathematical model. The calibrated asset models are subsequently used to either determine a suitable price for over-the-counter (OTC) exotic financial derivatives products, or for hedging
and risk management purposes. 

Calibrating financial models is a critical subtask within finance, and may need to be performed numerous times every day.
Relevant issues in this context include accuracy, speed and robustness of the calibration.
Real-time pricing and risk management require a fast and accurate calibration process. Repeatedly computing the values using mathematical models and at the same time fitting the parameters may be a computationally heavy burden, especially when dealing with multi-dimensional asset price models. The calibration problem is furthermore not necessarily a convex optimization problem, and it often gives rise to multiple local minima.

A generic, {\em robust} calibration framework may be based on a global optimization technique in combination with a highly efficient pricing method, in a parallel computing environment. To meet these requirements, we will employ the machine learning technology and develop an artificial neural network (ANN) solution method for a generic calibration framework.

 \emph{The proposed ANN-based framework comprises three phases, i.e., training, prediction and calibration.} During the training phase, the hidden layer parameters of the ANNs are optimized by means of supervised learning. This training phase builds a mapping between the model parameters and the output of interest. During the prediction phase, the hidden layers are kept unchanged (frozen) to compute the output quantities (e.g.,option prices) given various input parameters of the asset price model. The prediction phase can also be used to evaluate the model performance (namely testing). Together these steps are called the \emph{forward pass}. Finally, during the calibration phase, given the observed output data (e.g., market option prices), the original input layer becomes a learnable layer again, whereas all previously learned hidden layers are kept fixed. This latter stage is also called the \emph{backward pass}.  The overall calibration framework we name {\em CaNN (Calibration Neural Network)} here.  The CaNN inverts the already trained neural network conditional on certain known input.

There are several interesting aspects to the proposed approach. First of all, the machine learning approach may significantly accelerate classical option pricing techniques, particularly when involved asset price models are of interest. Recently there has been increasing interest in applying machine-learning techniques for fast pricing and calibration, see ~\citep{pricingNNs2019, Poggio2017,2018pricingwithgaussian, deeplearningvol:2019, 2018ModelUsingCNN,2016ModelUsingANN,DNNPricing2019}. For example, the paper ~\citep{2018pricingwithgaussian} used Gaussian process regression methods for derivative pricing. Other work, including this paper, employs artificial neural networks to learn the solution of the financial SDE system ~\citep{pricingNNs2019, deeplearningvol:2019, DNNPricing2019}, that do not suffer much from the curse of dimensionality. 

Secondly, there is inherent parallelism in our ANN approach, so we will also take advantage of modern processing units (like GPUs). The paper \citep{deeplearningvol:2019} also presented a neural network-based method to compute and calibrate rough volatility models. Our CaNN however incorporates a parallel global search method for calibration. Moreover, the CaNN is a generic ANN-based framework, and views the three phases, training/prediction/calibration, as a whole, the difference between them being just to change the learnable units. Furthermore, the proposed ANN approach can handle a flexible number of input market data. In other papers, like \citep{2016ModelUsingANN}, \citep{2018ModelUsingCNN}, the number of input data had to be fixed in order to fit the employed Convolutional Neural Networks.

Calibrating financial models often gives rise to non-convex optimization problems, for which local optimization algorithms may have convergence issues.
A local optimization technique is generally relatively cheap and fast, but a key factor is to choose an accurate initial guess.  
Otherwise it may fail to converge and get stuck in a local minimum. 
The authors in \citep{Gilli2012} vary two parameters of the Heston model (keeping the other parameters unchanged), and show that the objective function exhibits multiple local minima. Also in \citep{2011VolSurface} it is stated that multiple local minimal are common for calibration in the FX and commodities markets. 

To address robustness, global optimizers are popular to calibrate financial models, like the Differential Evolution (DE) technique, Particle Swarm optimization and Simulated Annealing, as their convergence does not rely on specific initial values. 
DE has been used to calibrate financial models \citep{2009DE4Heston, Gilli2012} and to train neural networks \citep{2008DE4ANN}.
Parallel computing may help to solve calibration problems with global optimization within reasonable time.

The contributions of this paper are three-fold.
First, we design a generic ANN-based framework for calibration. Apart from data generators, all the components and tasks are implemented on a unified computing platform. Second, a parallel global searcher is adopted based on a population-based optimization algorithm (here DE), an approach that fits well within the ANN-based calibration framework. Both the forward and backward passes run in parallel, tackling the computational bottleneck of global optimization and making the calibration time reasonable, even in the case of  employing a large neural network. Third, the key components are robust and stable: using a robust data generator in combination and the global optimization technique makes sure that the ANN-based calibration method does not get stuck in local minima.

The rest of the paper is organized as follows. In Section \ref{section:calibration}, the Heston and Bates stochastic volatility models and their calibration requirements are briefly introduced. In Section \ref{section:ann-caliration}, artificial neural networks are introduced as function approximators, in the context of parametric financial models. In this section, a generic machine learning framework for model calibration to find the global solution is presented as well. In Section \ref{section:experiments}, numerical experiments are presented to demonstrate the performance of the proposed calibration framework. Some details of the employed COS option pricing method are given in the appendix.

\section{Financial Model Calibration}
\label{section:calibration}
We start by  explaining  the stochastic models for the asset prices, the corresponding partial differential equations for the option valuation and the standard ways of calibrating these models. The open parameters in these models, that need to be calibrated with the help of an objective function are also discussed.
\subsection{Asset pricing models}
In the following subsections we present the financial asset pricing models that will be used in this paper, the Heston and Bates stochastic volatility models. European option contracts are used as the examples to derive the pricing models, however, other types of financial derivatives can be taken into consideration in a similar way.
\subsubsection{The Heston model}
One of the most popular stochastic volatility asset pricing models is the Heston model \citep{Heston1993AOptions}, for which the system of stochastic equations under the risk-neural measure $\mathbb Q$ reads, 
\begin{subequations} \label{eq:heston}
\begin{align}
& dS_t  = r S_t dt + \sqrt{\nu_t} S_t dW^s_t, \;\; S_{t_0}=S_0,\\
& d\nu_t = \kappa(\bar{\nu} - \nu_t) dt + \gamma \sqrt{\nu_t}dW^{\nu}_t, \;\; \nu_{t_0}=\nu_0, \\
& dW^s_t dW^{\nu}_t = \rho_{x,\nu} dt,
\end{align}
\end{subequations}
with $\nu_t$ the instantaneous variance, $r$ the risk-free interest rate and $W^s_t, W^{\nu}_t$ are two Wiener processes with correlation coefficient $\rho_{x,\nu}$ \footnote{For simplicity, $\rho \equiv \rho_{x,\nu}$ in the paper.}. 
To avoid negative volatilities, the asset's variance in Equation (\ref{eq:heston}) is modeled by a CIR process, which is proposed in \citep{CIR1985} to
model interest rates.
It precludes negative values for $\nu(t)$, so that when $\nu(t)$ reaches zero it subsequently becomes positive.
The process can be characterized as a mean reverting square-root process, 
with as the parameters $\bar{\nu}$ the long term variance, $\kappa$ the reversion speed; $\gamma$ is the volatility of the variance. An additional parameter is $\nu_0$, the $t_0$-value of the variance. 

By the martingale approach, the following two-dimensional Heston option pricing PDE is found,
\begin{eqnarray}
    \frac{\partial V}{\partial t}&+&rS\frac{\partial
V}{\partial S}+\kappa(\bar{\nu}-\nu_t)\frac{\partial V}{\partial \nu_t} 
+\frac12 \nu_t S^2\frac{\partial^2 V}{\partial S^2} \nonumber \\
    &+& \rho\gamma S
\nu_t \frac{\partial^2V}{\partial S\partial \nu_t}+\frac12\gamma^2
\nu_t\frac{\partial^2 V}{\partial \nu_t^2}-r{V}=0,
\label{eq:heston-pde}
\end{eqnarray}
with the given terminal condition $V(T,S,\nu;T,K)$, where $V=V(t,S,\nu;T,K)$ is the option price at time $t$.

\subsubsection{The Bates model}
Next to the Heston model, we will also consider its generalization, the Bates model~\citep{batesmodel}, by adding
jumps to the Heston stock price process. The model is described by the following system of SDEs:  
\begin{subequations} \label{eq:bates}
\begin{align}
& \frac{\displaystyle d S_t}{\displaystyle S_t}=\left(r-\lambda_{J}\mathbb E[e^J-1]\right)dt+\sqrt{\nu_t}dW^x_t+\left(e^J-1\right)d X^{\mathcal{P}}_t,\\
& d\nu_t = \kappa(\bar{\nu} - \nu_t) dt + \gamma \sqrt{\nu_t}dW^{\nu}_t, \;\; \nu_{t_0}=\nu_0, \\
& dW^s_t dW^{\nu}_t = \rho_{x,\nu} dt,
\end{align}
\end{subequations}
with $X_\mathcal{P}(t)$ a Poisson process with intensity $\lambda_{J}$, and $J$ being normally distributed jump sizes with expectation
$\mu_J$ and variance $\nu^2_J$, i.e.
$J\sim\mathcal{N}(\mu_J,\nu^2_J).$
The Poisson process $X_{\mathcal{P}}(t)$ is assumed to be independent of the Brownian motions and of the
jump sizes.
Clearly, we have three more parameters, $\lambda_{J}$, $\mu_J$ and $\nu^2_J$, to calibrate in this case. The corresponding option pricing equation is a so-called Partial Integro-Differential Equation (PIDE),
\begin{multline} 
\frac{\partial{V}}{\partial t} + \frac{1}{2}\nu_tS^{2}\frac{\partial^{2}{V}}{\partial S^{2}} + \rho\gamma \nu_t S\frac{\partial^{2}{V}}{\partial S\partial \nu_t} + \frac{1}{2}\gamma^{2}\nu_t\frac{\partial^{2}{V}}{\partial \nu_t^{2}} + (r-\frac{1}{2}\nu_t-\lambda_J(e^{\mu_J}-1))\frac{\partial{V}}{\partial S}
\\ +\kappa(\bar{\nu}-\nu_t)\frac{\partial{V}}{\partial \nu_t}  -(r+\lambda_J){V}+\lambda_J\int_{0}^{\infty}{V}P_{J}\left(x\right)dx=0,
\label{eq:Bates-pide}
\end{multline}
with the given terminal condition $V(T,S,\nu;T,K)
$, where $P_{J}\left(x\right)$ is the log-normal probability density
function of the jump magnitudes.

Both the Heston and Bates models do not give rise to analytic option value solutions and the governing P(I)DEs thus have to be solved numerically. There are several possibilities for this, like by means of finite difference PDE techniques, Monte Carlo, or numerical integration methods. We will employ a Fourier-type method, the COS method from \citep{Fang2009COS}, to obtain highly accurate option values, for the details we refer to the Appendix.
A prerequisite to using Fourier methods is the availability of the asset price's characteristic function. From the resulting option values, the corresponding Black-Scholes' implied volatilities will be determined by means of a robust root-finding iteration known as Brent's method \citep{brentmethod}.

\subsection{The calibration procedure}

Calibration refers to estimating the model parameters (i.e., the constant coefficients in the PDEs) given the samples of the market data. The market value of either option prices or implied volatilities,
with moneyness $m:=S_0/K$ and time to maturity $\tau:=T-t$, is denoted by $Q^*(\tau,m)$, and the corresponding model-based value is $Q(\tau,m; \Theta)$, with the parameter vector $\Theta \in \mathbb{R}^{n}$, where $n$ denotes the number of parameters to calibrate.
For the Heston model, $\Theta:=[\rho, \kappa,\gamma,\bar{\nu}, \nu_0]$, while for the Bates model we have, $\Theta:=[\rho, \kappa,\gamma,\bar{\nu}, \nu_0, \lambda_J, \mu_J, \sigma_J]$. 

The difference between the observed values and the ones given by the model is indicated by an error measure,
\begin{equation} \label{eq:error_def}
e_i := ||Q(\tau_i,m_i; \Theta) - Q^*(\tau_i,m_i)||, \; i=1,...,N
\end{equation}
where  $||\cdot||$ measures the distance, and $N$ is the number of available calibration instruments. The total difference is represented by the following target function,
\begin{equation} \label{eq:error_data}
  J(\Theta) := \sum _{i=1}^N\omega_i e_i + \bar\lambda ||\Theta||,
\end{equation} 
where $\omega_i$ are the corresponding weights and $\bar\lambda$ is a regularization parameter.  When $\omega_i=\frac{1}{N}$ and $\bar\lambda=0$ with squared errors in Equation (\ref{eq:error_data}), we obtain a well-known error measure, the MSE (Mean Squared Error).  When people wish to guarantee perfect calibration for ATM options (the options are most liquid in the market),  the corresponding weight value $\omega_i$ is sometimes increased. Usually calibrating financial models reduces to the following minimization problem, 
\begin{equation} \label{eq:min_calibration}
arg\min_{\Theta \in \mathbb{R}^{n}} J(\Theta), 
\end{equation}
which gives us a set of parameter values making the difference between the market and the model quantities as small as possible. 

The above formula is over-determined in the sense that $N>n$, i.e., the number of data samples is larger than the number of to-calibrate parameters. Equation (\ref{eq:min_calibration}) is usually solved iteratively to minimize the residual. Initially a set of parameter values is assigned and the corresponding model values are determined; These values are compared with market data, and the corresponding error is computed, after which a search direction is determined to find a next parameter set. The above steps are repeated until a stopping criterion is met. While evaluating Equation (\ref{eq:error_data}), an array of options with different strikes and maturities need to be valued thousands of times and therefore this valuation should be performed highly efficiently. 
\emph{Here, we will employ ANNs that can deal with a complete array of option prices in parallel.}

\subsection{Choices within calibration }
 Usually the objective function is highly nonlinear and even nonconvex. The authors in \citep{2012Calibration-Risk} discuss the impact of the objective function and the calibration method for the Heston model. 
This issue becomes worse when being faced with a high-dimensional optimization problem. A way to address this problem is to smooth the objective function and employ traditional local optimization methods. 
Another difficulty when calibrating the model is that the
set $\Theta$ includes multiple parameters that need to be
determined, and that these model parameters are not completely 
``independent'', for example, the effect of different parameters on the shape
of the implied volatility smile may be quite similar. For this reason, one may encounter several
``local minima'' when searching for optimal parameter values. In most cases, a global optimization algorithm should be preferred during calibration.

Regarding the target objective function, there are two popular choices in the financial context, namely either based on observed option prices or based on computed implied volatilities. Option prices can be collected directly from the market, and implied volatility should be computed based on the collected option prices.
 The most common choices without regularization terms include,
\begin{eqnarray}
\min_{\Theta} \sum_{i}\sum_{j} \omega_{i,j}\left(V_c^{*}(T_j-t_0,S_0/K_i) - V_c(T_j-t_0,S_0/K_i;\Theta)\right)^2,
\end{eqnarray}
and
\begin{eqnarray}
\min_{\Theta} \sum_{i}\sum_{j} \omega_{i,j}\left(\sigma_{imp}^{*}(T_j-t_0,S_0/K_i) - \sigma_{imp}(T_j-t_0,S_0/K_i;\Theta)\right)^2,
\end{eqnarray}
where $V_c^{*}(T_j-t_0,S_0/K_i)$ is the call option price for
strike $K_i$ and maturity $T_j$ with instantaneous stock price $S_0$ at time $t_0$ as observed in the market;
$V_c(T_j-t_0,S_0/K_i;\Theta)$ is the call option value computed from the model using model parameters $\Theta$; similarly
$\sigma^{*}_{imp}(\cdot)$, $\sigma_{imp}(\cdot)$ are the implied
volatilities from the market and from the Heston/Bates model,
respectively; $\omega_{i,j}$ is some weighting function. The notation $i$ and $j$ is to distinguish the two factors impacting the target quantity.    A third approach is to calibrate the model to both prices and implied volatility.  For option prices, weighting the target quantity by Vega (the derivative of the option price with respect to the volatility) is a technique to remedy model risk.   When taking implied volatility into account, a numerical root-finding method is often employed to invert the Black-Scholes formula in addition to computing option prices. That is to say, two numerical methods are required, one for pricing options, the other one for calculating the Black-Scholes implied volatility. Nevertheless calibrating to an implied volatility surface can help people specify prices of all vanilla options, given the current term structure of interest rates. This is one of the reasons why the practitioners prefer implied volatility during calibration. Besides, we will mathematically discuss the difference between calibrating to option prices and implied volatilities  in Section \ref{section:sensitivity analysis}.  Moreover, it is well known that OTM instruments are liquid or heavily traded in the market. Calibrating the financial models to OTM instruments is common practice in reality. 

The calibration performance (e.g., speed and accuracy) is also influenced by the employed method while solving the financial models. An analytic solution is not necessarily available for the model to be calibrated, and different numerical methods have therefore been developed to solve the corresponding option pricing models.  Alternatively, based on some existing solvers, ANNs can be used as a numerical method to learn the solution \citep{pricingNNs2019}.

\section{An ANNs-based approach to calibration} \label{section:ann-caliration}

This section presents the framework to calibrate a financial model by means of machine learning.
Training the ANNs and calibrating financial models both boil down to optimization problems, which motivates the present machine learning-based approach to model calibration. 

\subsection{Artificial Neural Networks} \label{section:ann formulas}

This section introduces the ANNs. In general, ANNs are built using three components: neurons, layers and the complete architecture from bottom to top. As the fundamental unit, a neuron consists of three consecutive operations, summing up the weighted input, adding a bias to the summation, and computing the output via an activation function. This activation function determines whether and by how much a particular neuron is active. A number of neurons make up a hidden layer. Stacking different layers then defines the full architecture of the ANNs. With signals travelling from the input layer through the hidden layers to the output layer, the ANN builds a mapping among input-output pairs. 

The basic ANN is the multi-layer perceptron (MLP), which can be written as a composite function,
\begin{equation} \label{eq:fun-dnn}
\mathnormal{ F(\mathbf{x}|\boldsymbol{\theta}) = f^{(L)}(...f^{(2)}(f^{(1)}(\mathbf{x};\boldsymbol{\theta}^{(1)});\boldsymbol{\theta}^{(2)});...\boldsymbol{\theta}^{(L)})},
\end{equation}
where $\boldsymbol{\theta}^{(i)}= (\mathbf{w}_i, \mathbf{b}_i)$ \footnote{Here, we use the notation $\theta\equiv \theta_{ANN}$, as compared to the parameters $\theta_{DE}$ of the DE, or $\theta_{Heston}$ of the Heston model.}, $\mathbf{w}_i$ is a weight matrix and $\mathbf{b}_i$ is a bias vector. A one hidden layer MLP can, for example, be written as follows, 

\begin{equation}\label{eq:nn-formula-one} 
\begin{cases} 
y(\mathbf{x})= \varphi^{(2)} \left(\sum_{j}w_{j}^{(2)}z^{(1)}_{j}+b^{(2)} \right) \\
z_{j}^{(1)} = \varphi^{(1)} \left(\sum_{i}w_{ij}^{(1)}x_{i}+b_{j}^{(1)} \right).
\end{cases}
\end{equation}
with $w_j$ the unknown weights, $\varphi(w_{1j}x_j+b_{1j})$ the neuron's basis function, $\varphi(\cdot)$ an activation function ($m$ is the number of neurons in a hidden layer). 

The loss function is equivalent to a distance in the case of supervised learning,
\begin{equation}\label{eq:lossfndef}
L(\boldsymbol{\theta}):=D(f(\mathbf{x}), F(\mathbf{x}|\boldsymbol{\theta})),
\end{equation}
where $f(\mathbf{x})$ is the target function. Training the ANNs is learning the optimal weights and biases in Equation (\ref{eq:fun-dnn}) to make the loss function as small as possible. The process of training neural networks can be formulated as an optimization problem,
\begin{equation} \label{eq:argmin_dnn}
arg\min_{\boldsymbol{\theta}} L(\boldsymbol{\theta} | (\mathbf{X},\mathbf{Y})),
\end{equation}
given the input-output pairs $(\mathbf{X},\mathbf{Y})$ and a user-defined loss function $L(\boldsymbol{\theta})$. Assuming the training data set $(\mathbf{X},\mathbf{Y})$ can define the true function on a domain $\Omega$, ANNs with sufficiently many neurons can approximate this function in a certain norm, e.g., the $l_2$-norm. 
ANNs are thus powerful universal function approximators and can be used without assuming any pre-specified relation between the input and the output. 

Quantitative theoretical error bounds for ANNs to approximate any function are not yet available. For continuous functions, in the case of a single hidden layer, the number of neurons should grow exponentially with the input dimensionality~\citep{NNsmoothfun1996}. In the case of two hidden layers, the number of neurons should grow polynomially. The authors in~\citep{LowerboundsMLP1999} proved that any continuous function defined on the unit hypercube $C[0,1]^d$ can be uniformly approximated to arbitrary precision by a two hidden layer MLP, with $3d$ and $6d+3$ neurons in the first and second hidden layer, respectively. In \citep{ErrorboundsRelu} the error bounds for approximating smooth functions by ANNs with adaptive depth architectures are presented. The theory gets complicated when the ANN structure goes deeper, however, these deep neural networks have recently significantly increased the power of ANNs, see, for example the Residual Neural Networks \citep{resnet_approximator2018}.

In order to perform the optimization in~(\ref{eq:argmin_dnn}), the above composite function~(\ref{eq:fun-dnn}) is differentiated using the chain rule. The first- and second-order partial derivatives of the loss function with respect to any weight $w$ (or bias $b$) are easily computable; for more details we refer to \citep{Goodfellow-et-al-2016}. This differentiation enables us to not only train ANNs with gradient-based methods, but also the sensitivity of the approximated functions using the trained ANN can be investigated. For this latter task, the Hessian matrix will be derived in Section \ref{section:experiments} to study the sensitivity of the objective function with respect to the calibrated parameters. 

\subsection{The forward pass: learning the solution with ANNs} \label{section: NNs4HestonBlack}
\begin{figure}[H]
  \centering
  \includegraphics[width=1.0\textwidth]{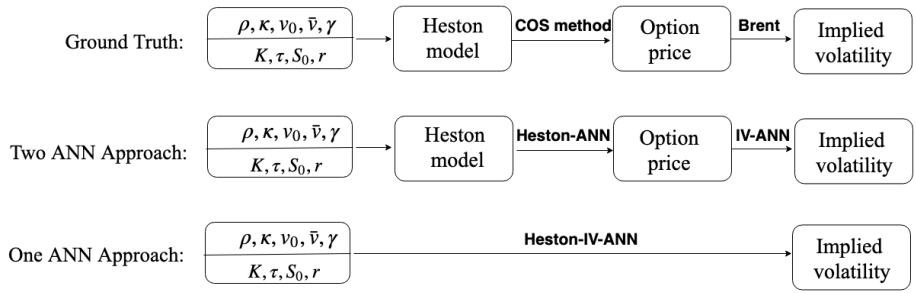}
  \caption{The Heston-implied volatility ANN.}
  \label{fig:flowchart Heston-IV-ANN solver}\label{figANNforwardpass}
\end{figure}

The first part of the CaNN, the forward pass, employs an ANN, in the form of an MLP,  to learn the solution generated by different numerical methods and subsequently maps the input to the output of interest (i.e., neglecting the intermediate variables). For example, in order to approximate the Black-Scholes implied volatilities based on the Heston input parameters, two numerical methods are required, i.e., the COS method to calculate the Heston option prices and Brent's root-finding algorithm to determine the corresponding implied volatility, as presented in Figure \ref{fig:flowchart Heston-IV-ANN solver}. Using two separate ANNs to map the Heston parameters to implied volatility has been applied in \cite{pricingNNs2019}. In the present paper, we merge these two ANNs, see Figure \ref{figANNforwardpass}. In other words, the Heston-IV-ANN is used as the forward pass to learn the mapping between the model parameters and the implied volatility. Note that a similar model is employed for the Bates model, however then based on the Bates model parameters. 

The forward pass consists of training and prediction, and in order to do so the network architecture and optimization method have to be defined. Generally, an increasing number of neurons, or a deeper structure, may lead to better approximations, but may also result in a computationally heavy optimization and evaluation of the network. 
In \citep{whydeepNNs} it is proved that a deep NN can approximate a function for which a shallow NN may need a very large number of neurons to reach the same accuracy.
Different residual neural networks have been trained and tested as a validation of our work. They may improve the predictive power while using a similar number of weights as in an MLP, but they typically take significantly more computing time during the training and testing phases. 
Very deep network structures may reduce the parallel efficiency, because the operations within a layer have to wait for the output of previous layers. With the limitation of computing resources available,  a trade-off between ANN's computation speed and approximation capacity  may be considered.

 Many techniques have been put forward to train ANNs, especially for deep networks. 
 Most of the neural network training relies on gradient-based methods. A proper {\em random initialization} may ensure the network to start with suitable initial weight values. {\em Batch normalization} scales the output of a layer by subtracting the batch mean and dividing it by the batch standard deviation. This can often speed up the training process.
A {\em dropout operation} randomly selects a proportion of the neurons and deactivates them, which forces the network to learn more generalized features and prevents over-fitting. The dropout rate $p$ refers to the proportion of deactivated neurons in a layer. In the testing phase, in order to take into account the missing activation during training, each activation in the entire network is reduced by a factor $p$. As a consequence, the ANNs prediction slows down, which has been verified during experiments on GPUs. We found that our ANNs model did not encounter over-fitting even when using a zero dropout rate, as long as sufficient training data were provided. In our neural network we employ the Stochastic Gradient Descent method, as further described in Section \ref{section:opt_algo}.

\subsection{The backward pass: calibration using ANNs}\label{section: NNs4HestonBlackBP}
This section discusses the connection between training the ANN and calibrating the financial model.  First of all, both Equations (\ref{eq:min_calibration}) and (\ref{eq:argmin_dnn}) aim at estimating a set of parameters to minimize a particular objective function. For the calibration problem, these are the parameters of the financial model and the objective function is the error measure between the market quantity and the model-based quantity. For the neural networks, the parameters correspond to the learnable weights and biases in the artificial neurons and the objective function is the user-defined loss. This connection forms an inspiration for the machine learning-based approach to calibrate financial models. 

\begin{figure}[H]
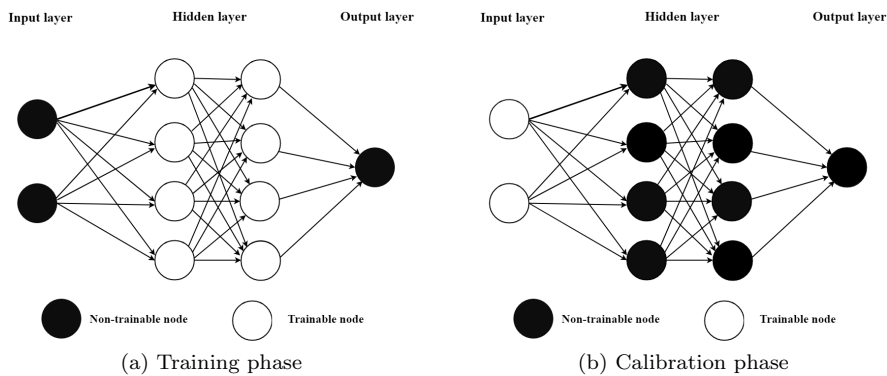
%
\centering
\subfloat[Training phase]{\label{fig:training phase} {\includegraphics[width=0.45\textwidth]{train_of_mlp} }}%
\qquad
\subfloat[Calibration phase ]{\label{fig:calibrating phase}{\includegraphics[width=0.45\textwidth]{calibration_of_mlp}}}%
\caption{The different phases of the ANNs.}%
\label{fig:ann_calibration}%
\end{figure}

As mentioned before, the ANN approach comprises three phases, training, prediction and calibration. During training, given the input-output pairs and a loss function as in Equation (\ref{eq:argmin_dnn}), the hidden layers are optimized to determine the appropriate values of the weights and biases, as shown in Figure \ref{fig:training phase}, which results in a trained ANN approximating the option solutions of the financial model (the forward pass, as explained in the previous section).

During the prediction phase, the hidden layers of the trained ANN are fixed (frozen), and new input parameters enter the ANN to yield the output quantities of interest. This phase is used to evaluate the performance of the trained ANN (the so-called model testing) or to accelerate option pricing by replacing the original solver.

During the calibration phase (or \emph{the backward pass}), the original input layer of the ANN is transformed into a learnable layer, while all hidden layers remain unchanged. These layers
are the ANN layers obtained from the forward pass with the already trained weights, as shown in Figure \ref{fig:calibrating phase}. By providing the output data, here consisting of market-observed option prices and implied volatilities, and changing to an objective function for model calibration, see Equation~(\ref{eq:min_calibration}), the ANN can be used to find the input values that match the given output. The task is thus to solve the inverse problem by learning a certain set of input values, here the model parameters $\Theta$, either for the Heston or Bates model.  The option's strike price $K$, as an example, belongs to the input layer, but is not estimated in this phase. 
Note that the training phase in the forward pass is time-consuming but done off-line and only once. The calibration phase is computationally cheap, and is performed on-line. The calibration phase thus results in model parameters that best match the observed market data. 

The gradients of the objective function, with respect to the input parameters, can be derived. This is useful when employing gradient-based optimization algorithms to conduct model calibration with the trained ANNs. Compared to the classical calibration methods, in the ANN-based approach it is also possible to incorporate the gradient information from the trained ANNs to compute the search direction (without external numerical techniques). As mentioned, we focus on a general calibration framework in which we integrate both gradient-based and gradient-free algorithms. Importantly, within the proposed calibration framework we may insert any number of market quotes, without requiring a fixed structure or a grid of input parameters.

\subsection{Optimization} \label{section:opt_algo}

 The optimization method plays a key role in training ANNs and calibrating financial models, but there are different requirements on the solutions for different phases. When training the neural network to learn the mapping between input and output values, we aim for a good performance on a test data set while optimizing the model on a training data set (this concept is called generalization). Calibration is regarded as an optimization problem with only a training data set, where the objective is to fit the market-observed prices as well as possible. In this work, the Stochastic Gradient Descent (SGD) is used when training the ANN, and Differential Evolution is preferred in the phase of calibration to address the problem of multiple local minima. \footnote{We have tested SGD during calibration. SGD is faster but may fail in some cases without good initial guess.}

\subsubsection{Stochastic Gradient Descent}
A popular optimizer to train ANNs is SGD \citep{SGDrobbins1951}. Neural networks contain thousands of weights, which gives rise to a high-dimensional, nonconvex optimization problem. The local minima appear not to be problematic for this involved black-box system, as long as the cost function reaches a sufficiently low value. Optimization of \eqref{eq:error_data} based on SGD is computed using,

\begin{equation}\label{eq:nn-formula}
\begin{cases} 
\mathbf{W}^{(i+1)} \leftarrow \mathbf{W}^{(i)} - \eta(i) \frac{\displaystyle \partial L}{\displaystyle \partial \mathbf{W}}, \\[1.5ex] 
\mathbf{b}^{(i+1)} \leftarrow \mathbf{b}^{(i)} - \eta(i) \frac{\displaystyle \partial L}{\displaystyle \partial \mathbf{b}},\\[1.5ex] \textnormal{for } i=0,1,...,N_T,
\end{cases}
\nonumber
\end{equation}
where $L$ is a loss function as in \eqref{eq:lossfndef} and $N_T$ is the number of training iterations. The bias and weights parameters are denoted by $\boldsymbol{\theta}=(\mathbf{W},\mathbf{b})$. The loss function of training the ANN solver is based on MSE in this paper.

In practice, the gradients are computed over mini-batches because of computer memory limitations. Instead of all input samples, a portion is randomly selected within each iteration to calculate an approximation of the gradient of the objective function. The size of the mini-batch is used to determine the portion. Due to the architecture of the GPUs, batch sizes of powers of two can be efficiently implemented. Several variants of SGD have been developed in the past decades, e.g., RMSprop and Adam, where the latter method handles an optimization problem adaptively by adjusting the involved parameters over time.

\subsubsection{Differential Evolution}

Differential Evolution (DE) \citep{DE1997} is a population-based, derivative-free optimization algorithm, which does not require any specific initialization. With DE, a global optimum can be found, even when the objective function is nonconvex. The general form of the DE algorithm usually comprises the following four steps:

\begin{enumerate}
\item Initialization: Generate the population with $N_p$ individuals and locate each member with random positions in the search space,
  $$ (\boldsymbol{\theta}_1, \boldsymbol{\theta}_2, ..., \boldsymbol{\theta}_{N_p}) $$
\item Mutation:
Once initialized, a randomly sampled difference is added to each individual, named differential mutation. 
\begin{equation}
  \boldsymbol{\theta'}_i = \boldsymbol{\theta}_a + F \cdot(\boldsymbol{\theta}_b- \boldsymbol{\theta}_c)
\end{equation}
where $i$ represents the $i$-th candidate, and the indices $a$, $b$, $c$ are randomly selected from the population with $a \neq i$. The resulting $\boldsymbol{\theta'}$ is called a mutant. The differential weight $F \in [0, \infty)$ determines the step size of the evolution. Generally, large $F$ values increase the search radius, but may cause DE to converge slowly. There are several mutation strategies, for example, when ${\theta}_a$ is always the best candidate of the previous population, the mutation strategy is called \emph{best1bin}, which will be used in the following numerical experiments; when ${\theta}_a$ is randomly chosen, it is called \emph{rand1bin}. After this step, an intermediary (or donor) population, consisting of $N_p$ mutant candidates, is generated.

\item Crossover:
During the crossover stage mutated candidates that may enter the next evaluation stage will be determined. For each ${ i\in \{1,\ldots,N_p\}}$, a uniformly distributed random number ${p_{i} \sim U(0,1)}$ is selected. Some samples are filtered out by setting a user-defined crossover possibility $Cr \in [0,1]$, 
\begin{equation}
 \centering 
 \boldsymbol{\theta}''_i=\Big\{
        \begin{array}{ll}
         \boldsymbol{\theta}'_i \text{, if } p_i \leq Cr,\\
         \boldsymbol{\theta}_i \text{, otherwise.} \\ 
         \end{array} 
\end{equation} 

If the probability is greater than $Cr$, the donor candidate will be discarded. Increasing $Cr$ allows more mutants to enter the next generation, but at the expense of population stability. Here, a trial population $(\boldsymbol{\theta}''_1, \boldsymbol{\theta}''_2, ..., \boldsymbol{\theta}''_{N_p})$ has been defined. 

\item Selection: Comparing each new trial candidate with the corresponding target individual on the objective function, 
\begin{equation}
 \centering 
 \boldsymbol{\theta}_{i} \leftarrow \Big\{
        \begin{array}{ll}
         \boldsymbol{\theta}''_i, \text{ if } g(\boldsymbol{\theta}''_i) \leq g(\boldsymbol{\theta}_i), \\
         \boldsymbol{\theta}_i, \text{ otherwise. } \\
         \end{array}
\end{equation}

If the trail individual has improved performance, the selected individual is replaced. Otherwise, the offspring individual inherits the parameters from its parent. This gives birth to a next generation population.

\end{enumerate}
The Steps (2)-(4) are repeated until the algorithm converges or until a pre-defined criterion is satisfied. Adjusting the control parameters may impact the performance of DE. For example, a large population size and mutation rate can increase the probability of finding the global minimum.  An additional parameter, convergence tolerance, is used to measure the diversity within a population, and determines when to stop DE. The control parameters can also change over time, which is out of our scope here. 

\subsubsection{Acceleration of calibration}
In this section we develop DE into a parallel version  which is beneficial within the ANNs. 
Generally, matrix multiplications and element-wise operations in a neural network can be implemented in parallel to reduce the computing time, especially when a large number of arguments is involved. As a result, several components of the calibration procedure can be accelerated. For the ANN solver in the forward pass, all observed market samples can be evaluated at once. Furthermore, in the selection stage of the DE, an entire population can be treated simultaneously. Note that the ANN solver runs in parallel, especially on any GPU. 

 \begin{table}[H]
  \caption{The setting of DE }
  \begin{center}
 \begin{tabular}{c | c }
 \hline
 Parameter & option \\
     \hline
Population size   & 50   \\
 Strategy  & best1bin   \\
Mutation   & (0.5, 1.0)    \\ 
Crossover recombination & 0.7  \\
Convergence tolerance    & 0.01  \\
\hline
\end{tabular}\label{table: DE setting}
\end{center}
\end{table}

An example of the parameter settings for DE is shown in Table~\ref{table: DE setting}, where the population of one generation comprises 50 vector candidates for the calibrated parameters (e.g., a vector candidate contains five parameters to calibrate in the Heston model), and each candidate produces a number of market samples (here 35, i.e., 7 strike prices $K$ and 5 time points).
 So, there are 50$\times$35=1650 input samples for the Heston model each generation. 
 Traditionally, all these input samples (here 1650) are computed individually, except for those with the same maturity time $T$. The first speed-up is achieved because 35 sample output quantities from each parameter candidate can be computed by the ANN solver at the same time, even if these samples have different maturity times and strike prices.
 The second speed-up is based on the parallel DE combined with the ANN, where all parameter candidates in one generation enter the ANN solver at once, that is, all 1650 input samples in one generation can be included in the ANN solver simultaneously,
 giving 1650 output values (e.g., implied volatilities). Note that the batch size of the ANN solver should be adapted to the limitations of the specific processor, here 2048 in our used processor. We find that with the population size being around 50, the parallel CaNN is at least 10 times faster than the conventional CaNN, on either a CPU or a GPU. It is believed that a larger population size should lead to a higher parallel computing performance, especially on a GPU.

\begin{remark}
There are basically two error sources in this framework. One is a consistency error which comes from the employed numerical methods to solve the financial model, and it is found while generating the training data set. The other is an optimization error during training and calibration. These errors will influence the performance of the CaNN.
\end{remark}

\section{Numerical results} \label{section:experiments}
In this section we show the performance of the proposed CaNN. We begin with calibrating the Heston model, a special case of the Bates model. Some insights into the effect of the Heston parameters on the implied volatility are discussed to give some intuition on the relation, since no explicit mapping between them exists. Then, the forward pass is presented where an ANN is trained to build a mapping between the model parameters and implied volatilities. It is also demonstrated that the trained forward pass can be used as a tool for performing the sensitivity analysis of the model parameters. After that, we implement the backward pass of the Heston-CaNN to calibrate the model and evaluate the CaNN performance. We end this section by considering the calibration of the Bates model, a model that consists of more parameters than the Heston model, using the Bates-CaNN.

\subsection{Parameter sensitivities for Heston model}
This section discusses the sensitivity of the implied volatility to the Heston coefficients. This sensitivity analysis can be used to estimate a set of initial parameters, as is used in traditional calibration methods. In our calibration method this will not be required, however, we can gain some insights in the case of no explicit formulas.

The typically observed implied volatility shapes in the market, e.g., the implied volatility smile or skew, can be reproduced by varying the above parameters $\{\kappa, \rho, \gamma, \nu_0, \bar{\nu}\}$. We will give some intuition about the parameter values and their impact on the implied volatility shape. From a PDE viewpoint, the calibration problem consists of finding appropriate values of PDE coefficients $\{\kappa, \rho, \gamma, \nu_0, \bar{\nu}\}$ to make the Heston model to reproduce the observed option/implied volatility data. 
The authors in \citep{2009SmartParameters4SABRandHeston} reduce the calibration time by giving smart initial values for asset models, whereas in \citep{2010AsymptoticHeston} an approximation formula for the Heston dynamics was employed to determine a satisfactory initial set of parameters, followed by a local optimization to reach the final parameters. The paper \citep{2017fastcalibration} derived a Heston model characteristic function to analytically obtain gradient information of the option prices during the search for an optimal solution. In Section \ref{section:sensitivity analysis} we will use the ANN to extract gradient information of the implied volatility with respect to the Heston parameters.

\subsubsection{Effect of individual parameters}

To analyze the parameter effects numerically, we use
 the following set of reference parameters, 
$$
T = 2, S_0   = 100, \kappa = 0.1, \gamma = 0.1, \bar{\nu}  = 0.1, \rho  = -0.75, \nu_0   = 0.05, r   = 0.05.
$$ 
A numerical study is performed by varying
individual parameters while keeping the others fixed. For each parameter set, Heston stochastic volatility option prices are computed (by means of the numerical solution of the Heston PDE) and 
the Black-Scholes implied volatilities are subsequently determined.

\begin{figure}[htb]
\centering
   \includegraphics[width=6.0cm]{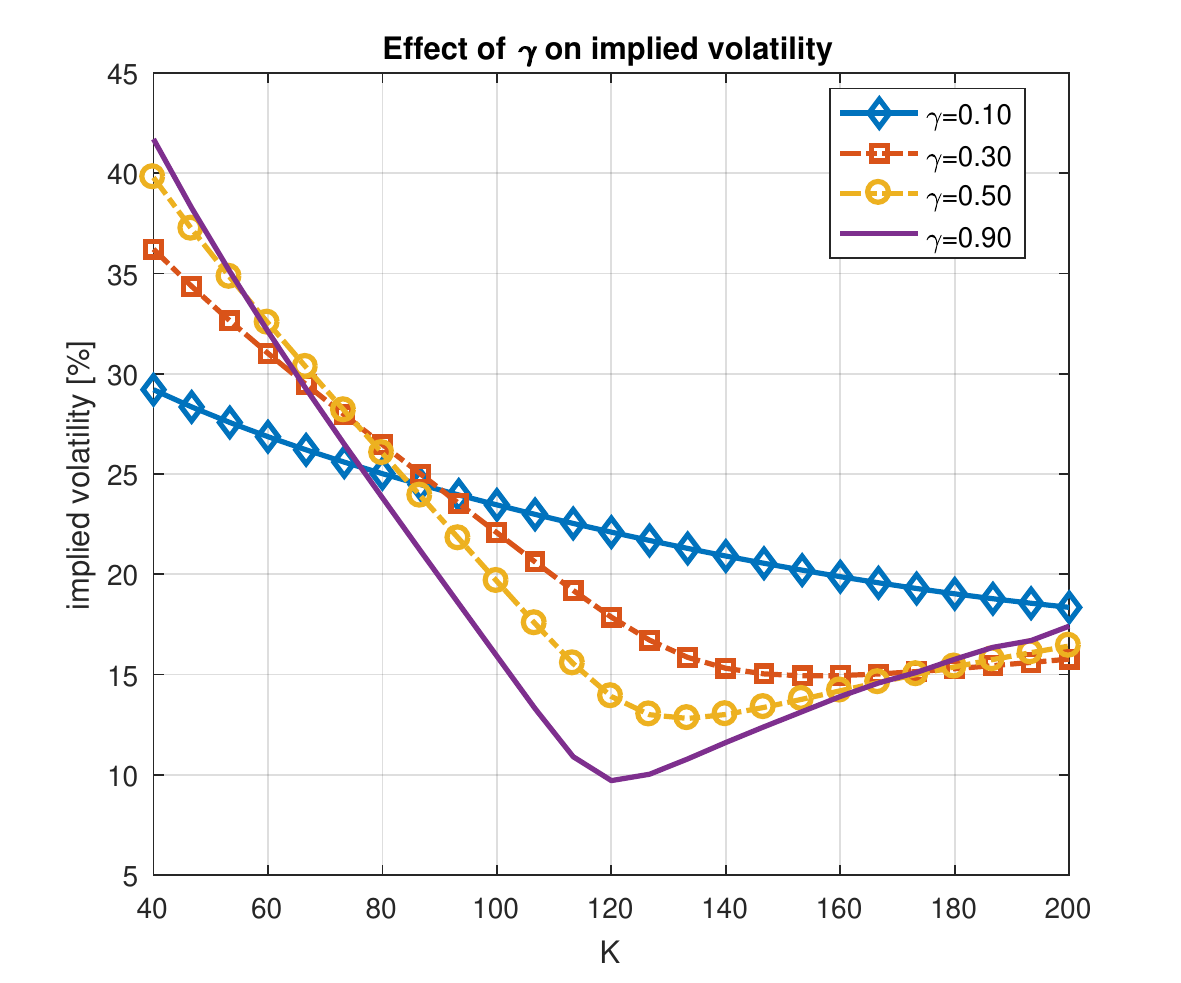}
   \includegraphics[width=6.0cm]{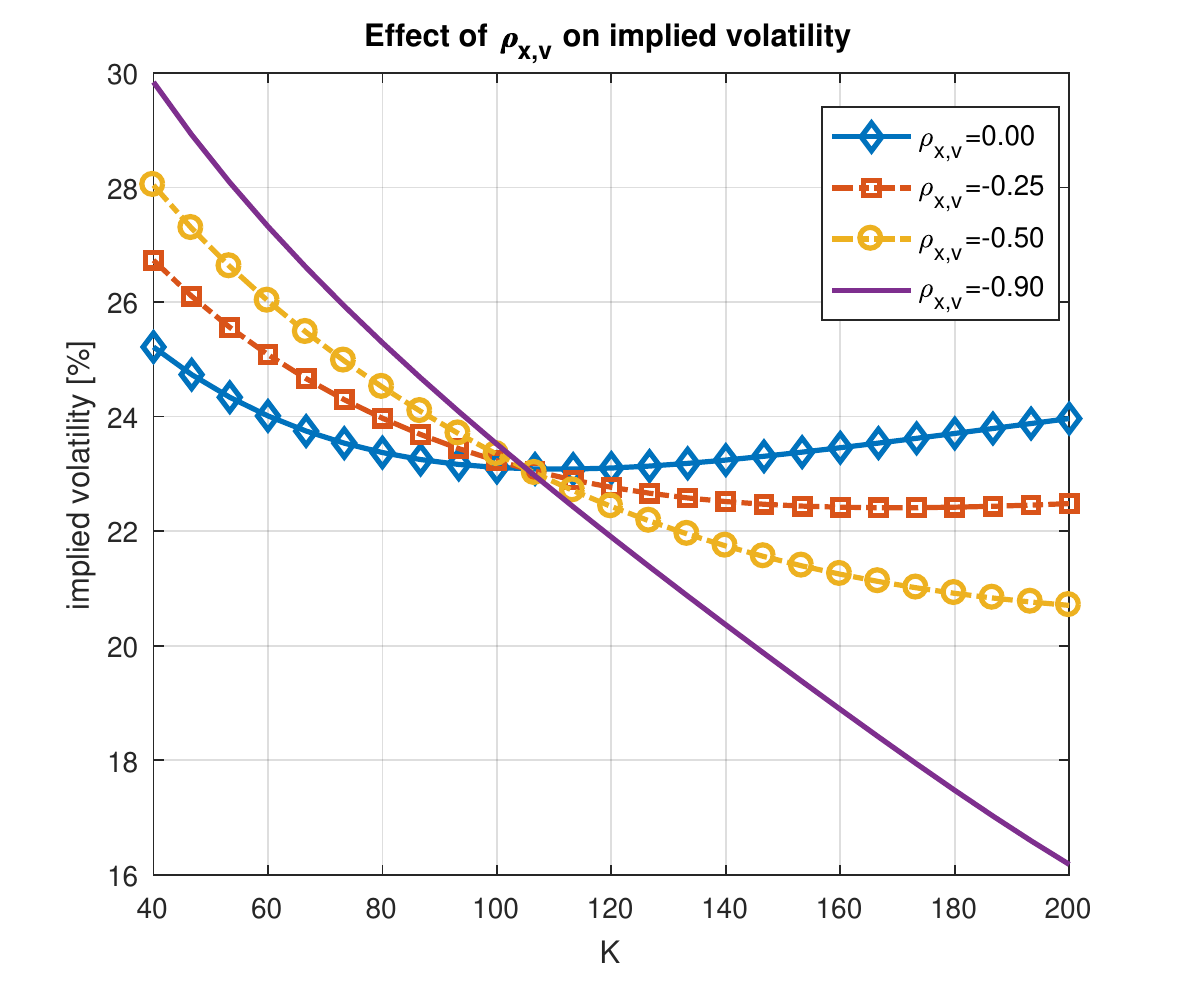}
	\caption{Impact of variation of the Heston parameter $\gamma$ (left side), and correlation parameter $\rho$ (right side), 
	on the implied volatility which varies as a function of strike price $K$. }
   \label{fig:ImpVol:1}
\end{figure}

Two important parameters that are varied are the correlation parameter
$\rho$ and the volatility-of-variance parameter $\gamma$.
	Figure~\ref{fig:ImpVol:1} (left side) shows that, when $\rho=0\%$, an increasing value of
$\gamma$ gives a more pronounced implied volatility {\it smile}. A higher volatility-of-variance parameter thus increases the implied volatility {\em curvature}.
	We also see, in Figure~\ref{fig:ImpVol:1} (right side), that when the
correlation between stock and variance process gets
increasingly negative, the slope of the {\em skew} in the implied volatility curve increases.
Furthermore, it is found that parameter $\kappa$ has a limited effect on
the implied volatility smile or skew, up to $1\%- 2\%$ only. It determines the speed at which the volatility converges to the long-term volatility $\bar{\nu}$.

The optimization can be accelerated by a reduction of the set of parameters to be optimized.
By comparing the impact of the speed of mean reversion parameter $\kappa$ and the
curvature parameter $\gamma$,
it is observed that these two parameters have a
similar effect on the shape of the implied volatility. It is
therefore common (industrial) practice to prescribe (or fix) one of them.
Practitioners often fix $\kappa=0.5$ and optimize parameter $\gamma$. By this,
the optimization reduces to four parameters.

Another parameter which may be determined in advance, using heuristics,
is the initial value of the variance process $\nu_0$.
For maturity time $T$ ``close to today'' (i.e., $T\rightarrow0$), one
expects the stock price to behave {\em like in the
Black-Scholes case}. The impact of a stochastic variance
process should reduce to zero, in the limit $T\rightarrow 0$.
For options with short maturities, the process may therefore be
approximated by a process
of the following form:
\begin{eqnarray}
d S(t) = r S(t)dt+ \sqrt{\nu_0}S(t)dW_x(t).
\end{eqnarray}
This suggests that for initial variance $\nu_0$ one may use the
square of the ATM implied volatility of an option with the
shortest maturity, $\nu_0\approx \sigma^2_{imp}$, for
$T\rightarrow0$, as an accurate approximation for the initial guess for the parameter. One may also use the connection of the Heston dynamics to the Black-Scholes dynamics
with a time-dependent volatility function.
In the Heston model we may, for example, {\em project} the variance process onto its expectation, i.e.,
\[d S(t)=rS(t)dt+\mathbb E[\sqrt{\nu(t)}] S(t)dW_x(t).\]
By this projection the parameters of the
variance process $\nu(t)$ may be calibrated similar to the case of the time-dependent
Black-Scholes model. 
The Heston parameters are then determined, such that
\[\sigma^{ATM}(T_i)=\sqrt{\int_0^{T_i}\left(\mathbb E[\sqrt{\nu(t)}]\right)^2dt},\]
where $\sigma^{ATM}(T_i)$ is the ATM implied volatility for
maturity $T_i$. 

Another classical calibration technique for the Heston parameters is to use VIX index market quotes. 
With different market quotes
for different strike prices $K_i$ and for different maturities $T_j$, we may
determine the optimal parameters by solving the following equalities, for all pairs
$(i,j)$,
\begin{eqnarray}
K_{i,j}=\bar{\nu}+\frac{\nu_0-\bar{\nu}}{\kappa(T_j-t_0)}\left(1-e^{-\kappa(T_i-t_0)}\right).
\end{eqnarray}
When the initial values of the parameters have been determined, one can use the whole
implied volatility surface to determine the
optimal model parameters. To conclude, the number of the Heston parameter to be calibrated depends on different scenarios. The flexibility of our CaNN is that it can handle varying numbers of to-calibrate parameters.

\subsubsection{Effect of two combined parameters} \label{section:losssurface}
\begin{figure}[H]
\subfloat[$\nu_0$ Vs $\kappa$]{\label{fig:v0Vskap_surface}{\includegraphics[width=0.5\textwidth]{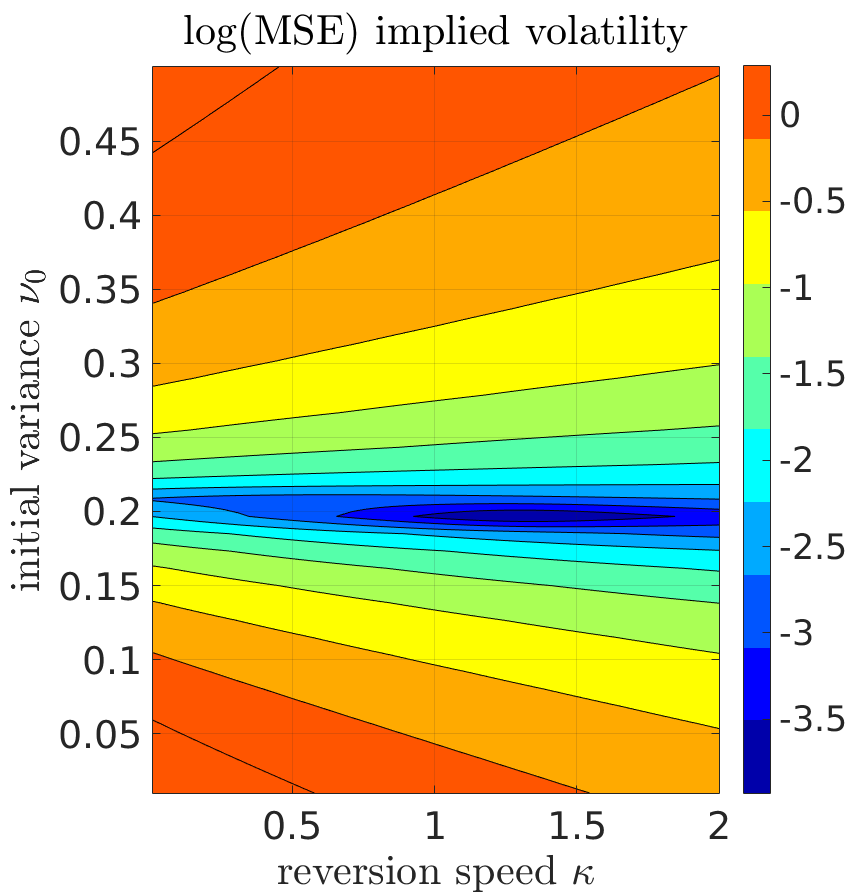}}}
\subfloat[$\bar{\nu}$ Vs $\kappa$]{\label{fig:vbarVskap_surface}{\includegraphics[width=0.5\textwidth]{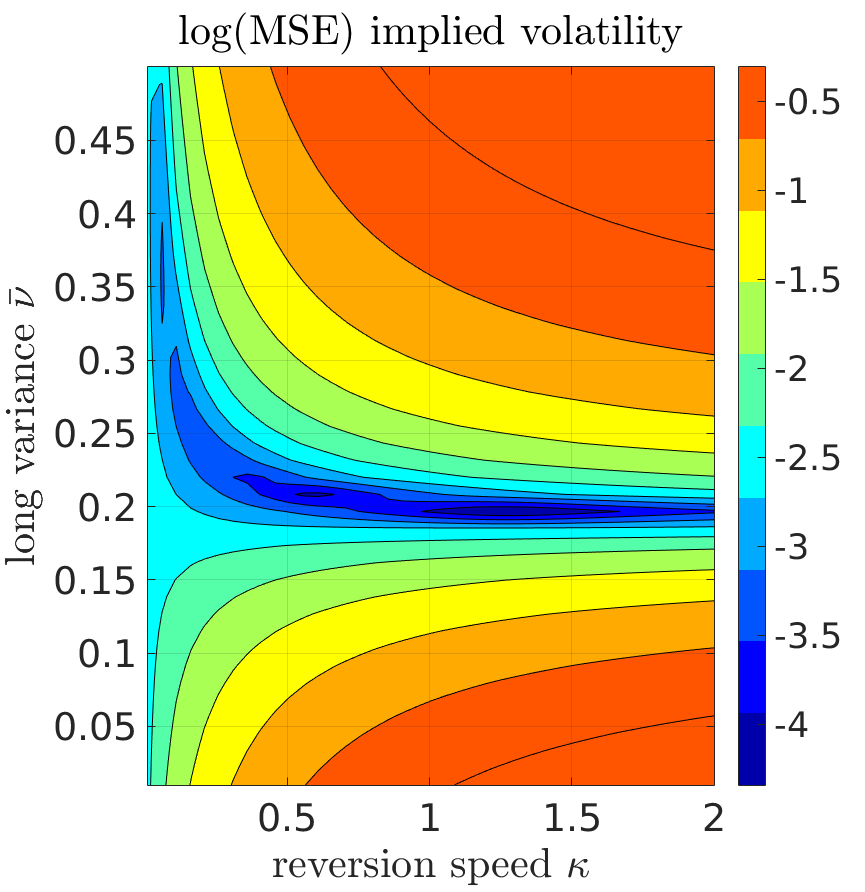}}} 
\caption{Landscape of the objective function for the implied volatility. The true values are $\kappa^*=1.0$ and $\nu_0^*=0.2$ in the left plot, and $\kappa^*=1.0$ and $\bar{\nu}^*=0.2$ in the right plot. There are 35 market samples. The objective function is MSE. The contour plot is rendered by a log-transformation. 
}
\label{fig:iv-loss-surface-kappa-v0}
\end{figure}

In this section, two parameters are varied simultaneously in order to understand the joint impact on the objective function.
Figure \ref{fig:v0Vskap_surface} presents the landscape of the objective function, here the logarithm of the MSE, when varying $\nu_0$ and $\kappa$ but keeping the other parameters fixed in the Heston model. It is observed that the valley is narrow in the direction of $\nu_0$ but flat in the direction of $\kappa$. Several values of these parameters thus result in similar values of the objective function, which means that there may be no unique global minimum above a certain error threshold. Furthermore, for $\bar\nu$ and $\kappa$ we observe also a flat minimum, with multiple local minima giving rise to similar MSEs, see Figure~\ref{fig:vbarVskap_surface}.

\begin{figure}[H]
 \centering
\includegraphics[width = 0.6\textwidth]{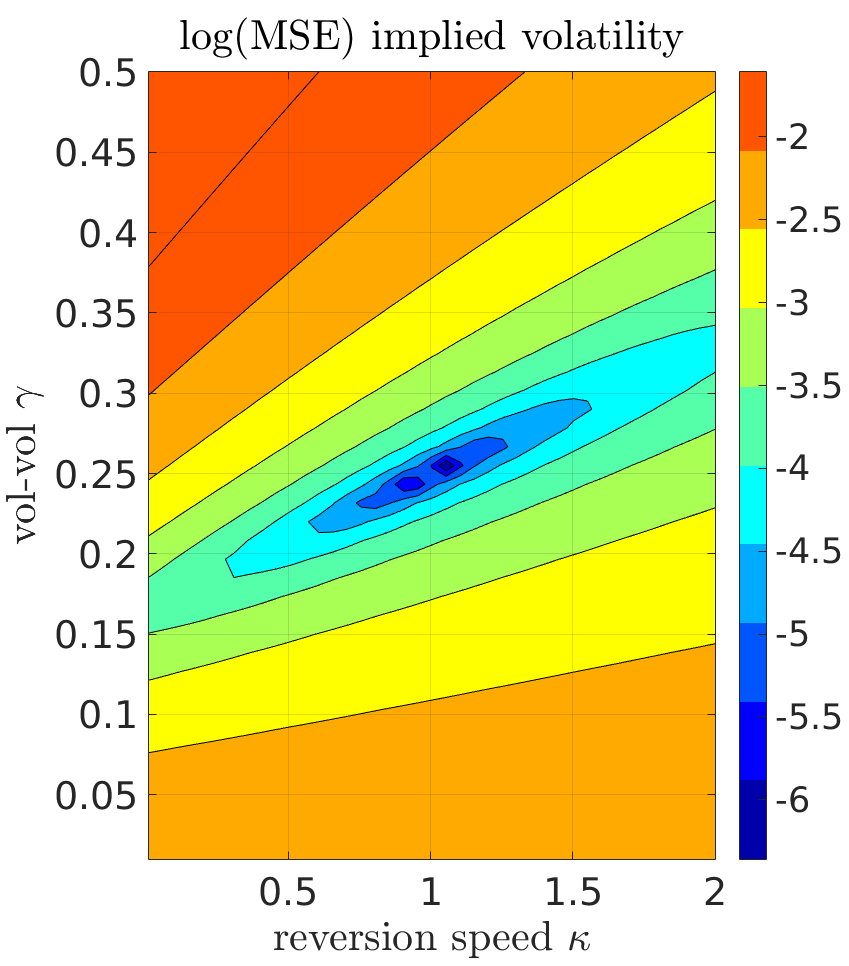}
\caption{The objective function when varying $\gamma$ and $\kappa$. The true values are $\kappa^*=1.0$ and $\gamma^*=0.25$.}
\label{fig:obj_fun kappa-gamma}
\end{figure}

A similar observation holds for $\kappa$ and $\gamma$: small values of $\kappa$ and large $\gamma$ values will, in certain settings, give essentially the same option prices as large values of $\kappa$ and small $\gamma$ values. This may give rise to multiple local minima for the objective function, as shown in Figure \ref{fig:obj_fun kappa-gamma}. 

For higher-dimensional objective functions, the structure becomes even more complex.
This is a preliminary study of the sensitives, and advanced tools are required for studying the effect of more than two parameters. We will show that the ANN can be used to obtain the sensitivities for more than two parameters to present the bigger picture of the dependencies and sensitivities. For this task the Hessian matrix of the five Heston parameters will be extracted (see Section \ref{section:sensitivity analysis}).

\subsection{The forward pass}
\begin{figure}[H]
\includegraphics[width = \textwidth]{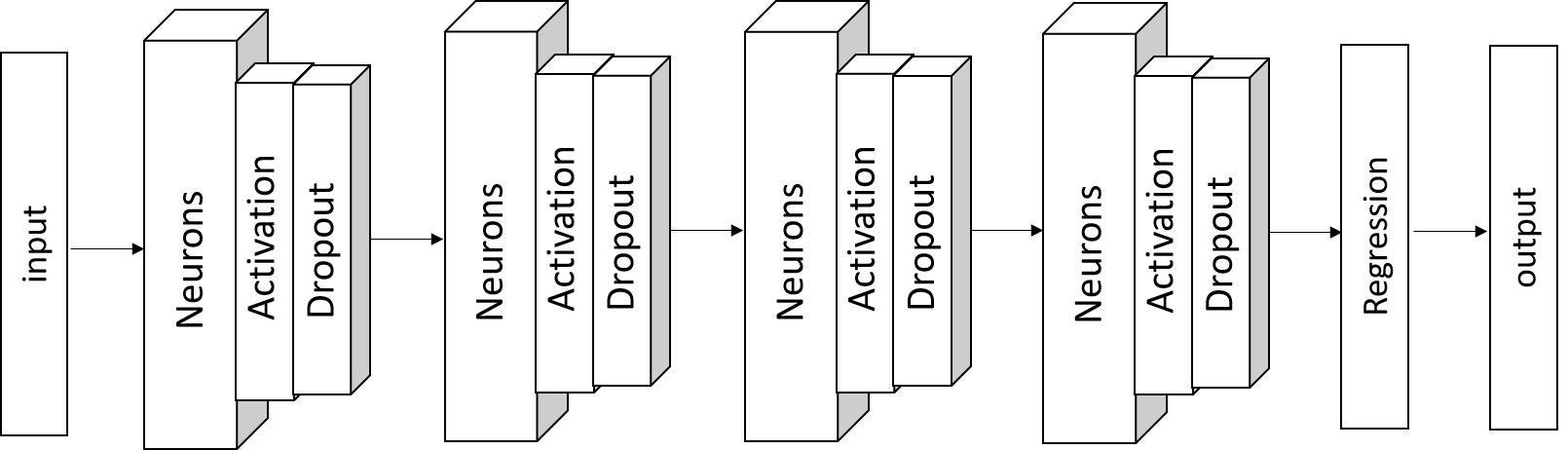}
\caption{The structure of the ANN.}
  \label{fig:pass}
\end{figure}

In this section, we discuss the forward pass, i.e., Heston-IV-ANN. The selected hyper-parameters are listed in Table \ref{table:final-nn-setting}. Increasing the number of neurons or using a deeper structure may lead to better approximations, but gives rise to an expensive-to-compute network. With our computing resources, we choose to employ 200 neurons each hidden layer to balance the calibration speed and accuracy. We use 4 hidden layers and a linear output (regression) layer, so that the network contains 122,601 trainable parameters.  MSE is used as the loss function measure to train the forward pass. 
The global structure is depicted in Figure~\ref{fig:pass}. More details on the ANN solver can be founded in \citep{pricingNNs2019}.

 \begin{table}[H]
\begin{center}
\caption{Details and parameters of the selected ANN.}
 \begin{tabular}{ c | c }
  \hline
  Parameters     & Options \\ \hline 
  Hidden layers & 4 \\
  \hline

  Neurons(each layer) & 200 \\

  Activation     & ReLu \\

  Dropout rate 		& 0.0 \\ 

  Batch-normalization & No \\

  Initialization   & Glorot\_uniform\\

  Optimizer      & Adam \\

  Batch size     & 1024 \\

  \hline	 

 \end{tabular}
  \label{table:final-nn-setting}
 \end{center}
 \end{table}

As a data-driven method, the samples from the parameter set for which the ANN is trained are randomly generated for the pricing of European put options. The input contains eight variables, and Table \ref{table:heston para-range-random-vol-version2} presents the range of six Heston input parameters ($r$, $\rho$, $\kappa$, $\bar{\nu}$, $\gamma$, $\nu_0$) as well as two option contract-related parameters ($\tau$, $m$), with a fixed strike price $K=1.$ There are around one million data points. The complete data set is randomly divided into three parts, with 10\% as the testing set, 10\% as validation and 80\% as the training data set. 

After sampling the parameters, a robust version of the COS method is used to determine the option prices under the Heston model numerically. The default setting with $L_{COS}=50$ and $N_{COS}=1500$ will provide highly accurate option solutions for most of the samples, but it may end up with insufficient precision in some extreme parameter cases. In such cases, the integration interval $[a, b]$ will be enlarged automatically, by increasing $L_{COS}$ until the lower bound $a$ and the upper bound $b$ have different signs. Subsequently, the Black-Scholes implied volatility is calculated by Brent's method.

The option prices are just intermediate variables during training in the forward pass. The overall Heston-IV-ANN solver does not depend on the type of European option (e.g., call or put), since during the computation of the Black-Scholes implied volatilities the European options with identical Heston parameters should give rise to
 the same implied volatilities, independent of call or put prices. The 
  forward pass can handle both call and put implied volatilities without requiring additional efforts.

 \begin{table}[H]
\begin{center}
 \caption{ Sampling range for the Heston parameters; LHS means Latin Hypercube Sampling, COS stands for the COS method (see the appendix), and Brent for the root-finding iteration. }
 \begin{tabular}{ c|c | c | c }
  \hline

  ANN & Parameters       & Value Range   & Generating Method \\ \hline

  \multirow{9}{*}{ANN Input} & Moneyness, $m=S_0/K$     & [0.6, 1.4] & LHS \\ 

  &Time to maturity, $\tau$    & [0.05, 3.0](year) & LHS \\

  &Risk free rate, $r$     & [0.0\%, 5\%] & LHS \\

  &Correlation, $\rho$     & [-0.90, 0.0] & LHS \\

  &Reversion speed, $\kappa$  &(0, 3.0] & LHS \\
  
  &Volatility of volatility, $\gamma$ &(0.01, 0.8] & LHS \\ 

  &Long average variance, $\bar{\nu}$ &(0.01, 0.5] & LHS\\

  &Initial variance, $\nu_0$  &(0.05, 0.5] & LHS\\ 

  \hline

  -    & European put price, $V$        & $(0, 0.6)$ & COS\\ 
  ANN Output& Black-Scholes IV, $\sigma$        & $(0, 0.76)$ & Brent\\ 

  \hline	 

 \end{tabular}
 \label{table:heston para-range-random-vol-version2}
 \end{center}
 \end{table}

 The ANN takes as input parameters ($r$, $\rho$, $\kappa$, $\bar{\nu}$, $\gamma$, $\nu_0$, $\tau$, $m$), and approximates the Black-Scholes implied volatility $\sigma$. As mentioned in Table \ref{table:final-nn-setting}, the  optimizer Adam is used to train the ANN on the generated data set.
 The learning rate is halved every 500 epochs.
The training consists of 8000 epochs, both the training and validation losses have converged. The performance of the trained model is shown in Table \ref{table:ANN-IV-Heston Performance}.
 
 \begin{table}[H]
\begin{center}
\caption{ The trained forward pass performance. The default float type is float32 on the GPU. }
\scalebox{1.0}{
 \begin{tabular}{ c | c | c | c | c  }
  \hline
     Heston-IV-ANN  & MSE & MAE & MAPE & $R^2$ \\ \hline
   Training &  $8.07\times 10^{-8}$  & $ 2.15\times 10^{-4}$& $ 5.83\times 10^{-4}$ & 0.9999936 \\
   Testing &  $1.23\times 10^{-7} $ & $2.40\times 10^{-4}$ & $7.20\times 10^{-4}$ & 0.9999903\\
  \hline	 
\end{tabular} }
\label{table:ANN-IV-Heston Performance}
\end{center}
\end{table}

We observe that the forward pass is able to obtain a very good accuracy and therefore learns the mapping between model parameters and implied volatility in a robust and accurate manner. The test set performance is very similar to the train performance, showing that the ANN is able to generalize well.

\subsection{The backward pass}
We will perform calibration using the CaNN based on the trained ANN from the previous section and evaluate its performance. We will work with the full set of Heston parameters to calibrate, but we will also study the impact of reducing the number of parameters to calibrate.

\begin{figure}[htp]
  \centering
  \includegraphics[width=1.0\textwidth]{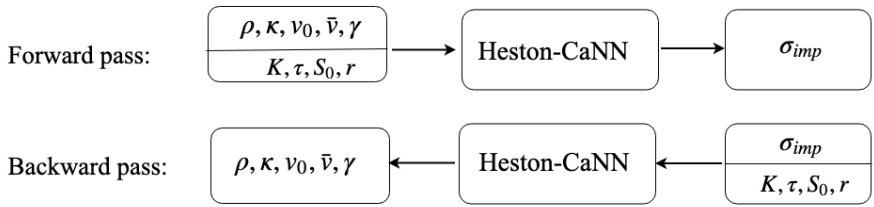}
  \caption{ The Calibration Neural Network for the Heston model.}
  \label{fig:Backward pass}
\end{figure}

 The aim is to check how accurately and efficiently the ANN approach can recover the input values. In order to investigate the performance of the proposed calibration approach, 
 as shown in Figure \ref{fig:Backward pass}, 
we generate synthetic samples by means of Heston-IV/COS-Brent, where the 'true' values of the parameters are known in advance. In other words, the parameters used to obtain the IV's from the COS-Brent's method, are now taken as output of the backward pass of the neural network, with $\sigma_{imp}$ being the input conditional on $(K,\tau, S_0, r)$. Different financial models correspond to different CaNNs. Here we distinguish the Heston-CaNN (based on the Heston model, studied in this section), from the Bates-CaNN (based on the Bates model, studied in Section \ref{section:bates-ann}).

There are $5\times7=35$ 'observed' European option prices, that are made up of European OTM puts and calls. As shown in Table \ref{table:Market parameter range},
the moneyness ranges from 0.85 to 1.15, and the maturity times vary from 0.5 to 2.0. 
Each implied volatility surface contains moneyness levels (85\%, 90\%, 95\%, 100\%, 110\%, 115\%) and maturities (0.5, 0.75, 1, 1.25, 1.5, 1.75, 2.0) with a prescribed risk-free interest rate of 3\%. The samples with $m<1$ correspond to European call OTM options, while those ones with $m>1$ and $m=1$ are OTM and ATM put options, respectively.

\begin{table}[H]
\begin{center}
\caption{The range of market quotes.}
 \begin{tabular}{ c| c|c | c  }
  \hline

 - &  Parameters         & Range   & Samples \\ \hline

  \multirow{4}{*}{ Market data}& Moneyness, $m=S_0/K$     & [0.85, 1.15] & 5 \\ 
  
  &Time to maturity, $\tau$    & [0.5, 2.0](year) & 7 \\

  &Risk free rate, $r$     & 0.03 & Fixed \\
  
  & European call/put price, $V/K$        & (0.0, 0.6) & -\\
  \hline
 Black-Scholes & Implied Volatility        & (0.2, 0.5) & 35\\

  \hline	 

 \end{tabular}
 \label{table:Market parameter range}
 \end{center}
 \end{table}

We use the total squared error measure $J(\Theta)$ as the objective function during the calibration,
\begin{equation} \label{eq:obj}
J(\Theta) = \sum \omega (\sigma^{\text{ANN}}_{imp} - \sigma_{imp}^*)^2 + \bar\lambda ||\Theta||,
\end{equation}
where $\sigma^\text{{ANN}}_{imp}$ is the ANN-model-based value and $\sigma_{imp}^*$ is the observed one. We give a small penalty parameter $\bar\lambda$ depending on the dimensionality of the calibration\footnote{When calibrating three parameters, we set $\bar\lambda$ to zero. When calibrating more than three parameters, $\bar\lambda$ is a small value $1.0\times 10^{-6}$, which is close to the MSE of the trained ANN. In this case, the regularization term only has a limited effect on the objective function during calibration.}. 
The forward pass has been trained with implied volatility as the output quantity, as described in Section \ref{section:losssurface}. The parameter settings of the DE optimization is shown in Table \ref{table: DE setting}.

\subsubsection{Calibration to Heston option quotes}
In this section we focus on two scenarios for the Heston model, calibrating either three parameters, with a fixed $\kappa$ and a known $\nu_0$, 
or calibrating five parameters. In order to create synthetic calibration data, we choose five equally-spaced points between the lower and upper bound for each parameter, and there are $5^5=3125$ combination cases in total, as shown in Table \ref{table:synthetic cases for calibration}. For each experiment, five different random seeds of DE are tested, because the DE optimization involves random operations which may cause the performance to fluctuate. In addition, all quotes have the equal weight $\omega=1$ in this section.

First, the scenario of three parameters is studied, fixing $\kappa$ and $\nu_0$ during calibration. We compare the averaged results by implementing each test case five times.
The wording ``function evaluation'' refers to how many times the model has been compared to the observed implied volatility. The population size in the DE is $15 \times N_v$, that is, $15\times3=45$. With the population ratio increasing further, no significant benefits were observed. 
As shown in Table \ref{table:3para-CaNN}, the time on the GPU is around half of that on the CPU.

\begin{table}[H]
\caption{Uniformly distributed points between the lower and upper bounds of the Heston parameters.}
\begin{center}
\begin{tabular}{ c | c c c | c}
\hline
parameter  & lower & upper & points & CaNN search space \\
\hline
$\rho$     & -0.75  & -0.25 & 5 &[-0.85,-0.05] \\
$\bar{\nu}$   & 0.15  & 0.35  & 5 &[0.05, 0.45] \\
$\gamma$    & 0.3   & 0.5  & 5  &[0.05, 0.75] \\
\hline
$\nu_0$     & 0.15  & 0.35 & 5 &[0.05, 0.45] \\
$\kappa$   & 0.5   & 1.0  & 5   &[0.1, 2.0] \\
\hline
\end{tabular}
\label{table:synthetic cases for calibration}
\end{center}
\end{table}

 \begin{table}[H]
 \caption{Averaged performance of the backward pass of the Heston-CaNN, calibrating $\mathbf{3}$ parameters on a CPU (Intel i5, 3.33GHz with cache size
4MB) and on a GPU (NVIDIA Tesla P100), over 3125$\times$5 (random seeds) test cases.}
 \begin{center}\footnotesize
 \renewcommand{\arraystretch}{1.3}
 \begin{tabular}{ lr | ll | lr}\hline\hline
 \multicolumn{2}{c}{ Absolute deviation from $\bm{\theta}^*$} & \multicolumn{2}{c}{Error measure}  & \multicolumn{2}{c}{Computational cost}\\ \hline
 $|\bar{v}^\dag-\bar{v}^*|$     & $1.60\times 10^{-3}$
  & $J(\Theta)$     & $1.45\times 10^{-6}$  & \textbf{CPU} time (seconds) & $0.29$ \\
 $|\gamma^\dag-\gamma^*|$     & $1.79\times 10^{-2}$ 
   & MJ & $4.14\times 10^{-8}$ & \textbf{GPU} time (seconds) & $0.15$ \\
 $|\rho^\dag-\rho^*|$   & $2.44\times 10^{-2}$     & Data points & 35 & Function evaluations & $59221$ \\

 \hline\hline
 \end{tabular}\label{table:3para-CaNN}
 \end{center}
 \end{table}

In the case of five parameters $(\rho,\bar{\nu},\gamma,\nu_0, \kappa)$, the calibration problem is more likely to give rise to a many-to-one problem; that is, many sets of parameter values may correspond to the same volatility surface. A regularization factor $\bar\lambda = 1.0\times10^{-6}$ is added to guide CaNN to a set of values for which the sum of their magnitude is the smallest among the feasible solutions, as shown in Equation (\ref{eq:error_data}). Here the DE population size is $50 = 10 \times 5$ parameters. As shown in Table \ref{table:callputotm 5para GPU parallel DE callputotm}, the Heston-CaNN finds the values of these parameters in approximately 0.5 seconds on a GPU, with around 20,000 function evaluations.
 \emph{There are several reasons why the CaNN with DE performs fast and efficiently.} One reason is that the forward pass runs faster compared to a two-step computation from the Heston parameters to the implied volatilities, since an iterative root-finding algorithm for the implied volatility takes some computing time. In addition, the entire group of observed data can be evaluated at once in the  framework. Other benefits come from the acceleration due to the parallelized DE optimization, where the whole population is computed simultaneously in the selection stage. 
 
 \begin{table}[H]
 \caption{Performance of Heston-CaNN, calibrating $\mathbf{5}$ parameters on a GPU, over 3125$\times$5 (random seeds) test cases. }
 \begin{center}\footnotesize
 \renewcommand{\arraystretch}{1.3}
 \begin{tabular}{ lr | lr | lr}\hline\hline
 \multicolumn{2}{c}{\emph{Absolute} deviation from $\bm{\theta}^*$} & \multicolumn{2}{c}{Error measure}  & \multicolumn{2}{c}{Computational cost}\\ \hline
  $|\nu_0^\dag-\nu_0^*|$     & $4.39\times 10^{-4}$   & $J(\Theta)$    & $2.52\times 10^{-6}$  & CPU time (seconds) & $0.85$ \\
 
 $|\bar{\nu}^\dag-\bar{\nu}^*|$     & $4.54\times 10^{-3}$    & MJ & $7.18\times 10^{-8}$ & GPU time (seconds) & $0.48$ \\
 
 $|\gamma^\dag-\gamma^*|$     & $ 3.28\times 10^{-2}$    & & & Function evaluations & $193249$ \\
 
 $|\rho^\dag-\rho^*|$   & $4.84\times 10^{-2}$      & & & Data points & $35$ \\
 $|\kappa^\dag-\kappa^*|$     & $4.88\times 10^{-2}$   & & & \\
 
 \hline\hline
 \end{tabular}\label{table:callputotm 5para GPU parallel DE callputotm}
 \end{center}
 \end{table}

\subsubsection{Sensitivity analysis based on ANNs} \label{section:sensitivity analysis}

The gradients of the objective function can be extracted from the trained model, as mentioned in Section \ref{section:ann formulas}. These can be used to gain some insights into the
complex structure of the loss surface and thus into the complexity of the optimization problem for calibration. We use here the Hessian matrix, which describes the local curvature of the loss function. No explicit formula is available for the relations the neural network learns between the implied volatilities and the model parameters, however, it is feasible to extract the Hessian from the trained ANN, giving insight into this relation and the sensitivities. Table \ref{table:Hessian} shows a Hessian matrix, where the Hessian is defined as $\partial_{y_iy_j}L(\bold{\theta})$, where $y_i$ and $y_j$ are output of the neural network (the to-be calibrated parameters). The Hessian is computed by differentiating the Heston-IV-ANN loss for computing the Black-Scholes implied volatility with respect to the Heston parameters on 35 market data points based on the parameter ranges in Table \ref{table:Market parameter range}. Here the objective function is the MSE to exclude the effects of a regularization factor. 

\begin{table}[H]
\begin{center}
 \caption{ A Hessian matrix at the true value set $\Theta^*$.}
\begin{tabular}{c|ccccc}
\hline
{} & $\partial \rho$ & $\partial \kappa$ & $\partial \gamma$ & $\partial \bar{\nu}$ & $\partial \nu_0$ \\
\hline

$\partial \rho$    &     2.79$\times 10^{-2}$ &      -- &     -- &        -- &     -- \\
$\partial \kappa$   &     1.14$\times 10^{-2}$&      8.20$\times 10^{-3}$ &     -- &        -- &     -- \\
$\partial \gamma$   &    -2.88$\times 10^{-2}$ &     -1.76$\times 10^{-2}$ &      4.11$\times 10^{-2}$ &       -- &     -- \\
$\partial \bar{\nu}$ &     7.45$\times 10^{-2}$ &      5.51$\times 10^{-2}$ &     -1.19$\times 10^{-1}$ &        3.76$\times 10^{-1}$ &     -- \\
$\partial \nu_0$   &     2.16$\times 10^{-1}$ &      1.27$\times 10^{-1}$ &     -3.10$\times 10^{-1}$ &        8.77$\times 10^{-1}$ &     2.66 \\
\hline
\end{tabular}

\label{table:Hessian}
 \end{center}
 \end{table}

 We can understand how the parameters affect the loss surface around the optimum with help of the Hessian matrix, by analyzing the sensitivities of the implied volatility with respect to the five parameters.
Observe that the value of the Hessian with respect to $\kappa$ is the smallest among the sensitivities. As shown in Table \ref{table:Hessian}, the ratio between $\partial^2 J(\Theta^*)/\partial \nu_0^2$ and $\partial^2 J(\Theta^*)/\partial \kappa^2$ is around 323, which suggests that changing 1 unit of $\nu_0$ is approximately equivalent to changing 323 units of $\kappa$ for the objective function. When the Hessian value is small in absolute value, the loss surface at that point exhibits flatness in the corresponding direction. As visible in Figure \ref{fig:iv-loss-surface-kappa-v0}, the ground-truth loss surface gets increasingly stretched along the axis with $\kappa$, resulting in a narrow valley with a flat bottom. This also indicates that there is no unique global minimum above a certain non-zero convergence tolerance, since multiple values of $\kappa$ would result in similar values of the loss function.
In addition, the convergence performance, especially for the steepest descent method, depends on the ratio of the smallest to the largest eigenvalue of the Hessian; this ratio is also known as the condition number in the case of symmetric positive matrices. The ratio between $\partial^2 J(\Theta^*)/\partial \bar{\nu}^2$ and $\partial^2 J(\Theta^*)/\partial \kappa^2$ is around 45, as visible in Figure \ref{fig:vbarVskap_surface}. From the results in \citep{2017fastcalibration}, when the target quantity is based on the option prices, this ratio between $\partial^2 J(\Theta^*)/\partial \bar{\nu}^2$ and $\partial^2 J(\Theta^*)/\partial \kappa^2$ is sometimes found to be of order $10^6$, which makes the calibration problem increasingly complex due to a great disparity in sensitivity. Calibrating to the implied volatility appears to reduce the ratio between different Hessian entries compared to the option prices, thus decreasing Hessian's condition number and resulting in a more efficient and accurate calibration performance.

Table \ref{table:Hessian} also suggests that 
the entries $|\partial^2 J(\Theta^*)/\partial \kappa^2|$ and $|\partial^2 J(\Theta^*)/\partial \rho^2|$ are among the smallest ones around the optimum. These two parameters thus have the smallest effect on the objective function. Therefore, the DE method can converge to values that are in a wide area of the search space, since these parameters do not impact the error measure significantly. A straightforward way to address this issue is  by adding a regularization term to choose a particular solution, for example,  like Equation (\ref{eq:obj}). Another way is to take advantage of the population-based algorithm DE. Since there are several candidates in each generation, we can select the top few candidates to get an averaged solution when DE converges. This averaged solution may lead to wider optima and better robustness. Some recent papers, like~\citep{2018SWA} have used similar ideas to improve the generalization of the neural network.   The parameter $\nu_0$ is the most sensitive one and it appears to dominate the ANN calibration process. Therefore,  the predicted parameter $\nu_0$ is the most precise among all parameters in order to achieve the desired accuracy.

The above analysis explains the behavior of the absolute deviation of the five parameters as shown in Table \ref{table:callputotm 5para GPU parallel DE callputotm}. The error measure MJ can not drop significantly below $7.18\times10^{-8}$, as this value is close to the testing accuracy, MSE$=1.23\times10^{-7}$, of the Heston-IV-ANN model. In other words, any further exploration of the DE optimization can not distinguish the parameters impact on the loss anymore.

\subsubsection{Calibration to Bates quotes} \label{section:bates-ann}

In this section, we use the Bates model to create the synthetic market data, in order to generate a more realistic (complex) volatility shape by adding some 'perturbations' to the previous Heston data. It is then followed by a calibration based on the Heston model. The aim is to check whether the resulting implied volatilities can be recovered by the machine learning calibration framework.

\begin{table}[H]

\caption{The Heston parameters are estimated with the CaNN by calibrating to a data set generated by the Bates model.  'Ground total squared error' refers to the sum of the differences between $\sigma^*_{imp}$ and $\sigma_{imp}$, where $\sigma_{imp}$ is obtained using the COS and Brent methods with already calibrated Heston parameter values. For a single calibration case, the computing time fluctuates slightly, as the CPU or GPU performance may be influenced by external factors. Function evaluations should be a reliable measure to estimate the time.}  
\makebox[1\textwidth]{
\scalebox{0.8}
{
 \begin{tabular}{ c|c | c c | c c | c c }

  \hline
     -- & Calibration  & Rare & Jump & Common & Jump & Weighting & ATM\\ \hline
  Parameters & Search space & Bates & Heston & Bates & Heston & Bates & Heston \\ \hline
  
  Intensity of jumps, $\lambda_J$ & - & 0.1 & - & 1.0 & - & 1.0 & - \\ 
  
  Mean of jumps, $\mu_J$    & - & 0.1 & - & 0.1 & - & 0.1 & -\\

  Variance of jumps, $\nu^2_J$ & -  & $0.1^2$ & - & $0.1^2$ & - & $0.1^2$ & - \\
 \hline

  Correlation, $\rho$ & [-0.9, 0.0]  &-0.3 & -0.284 &-0.3 & -0.135 &-0.3 & -0.164 \\

  Reversion speed, $\kappa$ & [0.1, 3.0] & 1.0 & 1.140 & 1.0 & 1.050 & 1.0 & 1.205 \\

  Long variance, $\bar{\nu}$ & [0.01, 0.5]  & 0.1 & 0.100 & 0.1 & 0.120 & 0.1 & 0.114\\

  Volatility of volatility, $\gamma$ & [0.01, 0.8]  & 0.7 & 0.728 & 0.7 & 0.701 & 0.7 & 0.604 \\  

  Initial variance, $\nu_0$  & [0.01, 0.5]  & 0.1  & 0.103 & 0.1  & 0.119 & 0.1 & 0.115 \\ 
  \hline
  \hline
  Function evaluations & CaNN& - & 162890 & - & 155680 & - & 258300\\
  Time(seconds)    & GPU & - & 0.45 & - & 0.40 & - & 0.7 \\
    \hline   
  Total Squared Error & Ground& - & 1.38$\times10^{-6}$ & - & 5.19$\times10^{-6}$ & - & 5.95$\times10^{-5}$\\ 
  \hline
 \end{tabular}
 }
 \label{table:Heston-Bates parameter}
 }
 \end{table}
 
 So, the observed data set in Table \ref{table:Market parameter range} is from the Bates option prices. During the calibration, we will employ the backward pass based on the Heston model to determine a set of parameter values which approximate the generated implied volatility function.

\begin{figure}[H]%
\centering
\subfloat[rare jump, equal weighting]{{\includegraphics[width=0.5\textwidth]{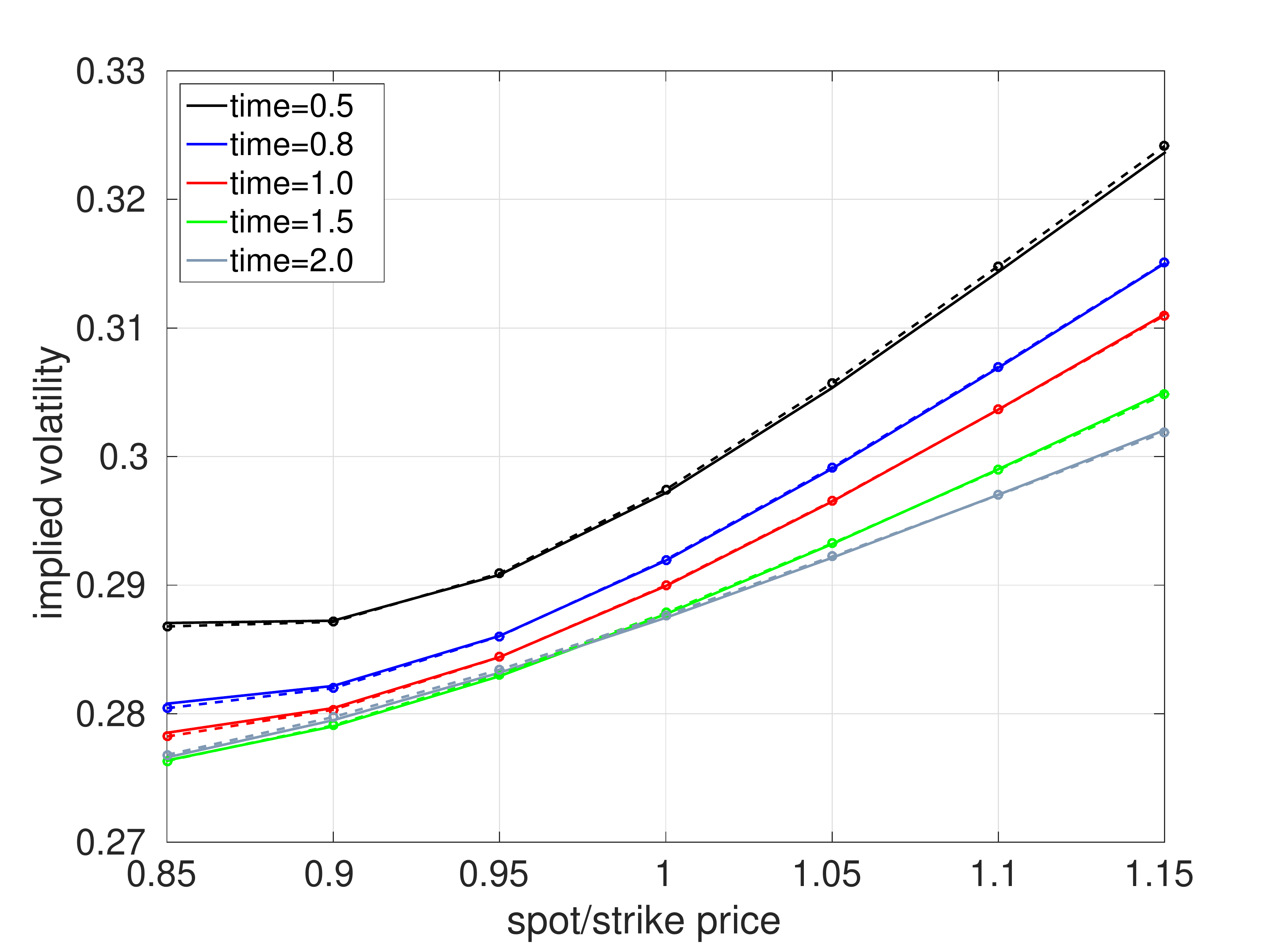} }}
\\
\subfloat[common jump, equal weighting]{{\includegraphics[width=0.5\textwidth]{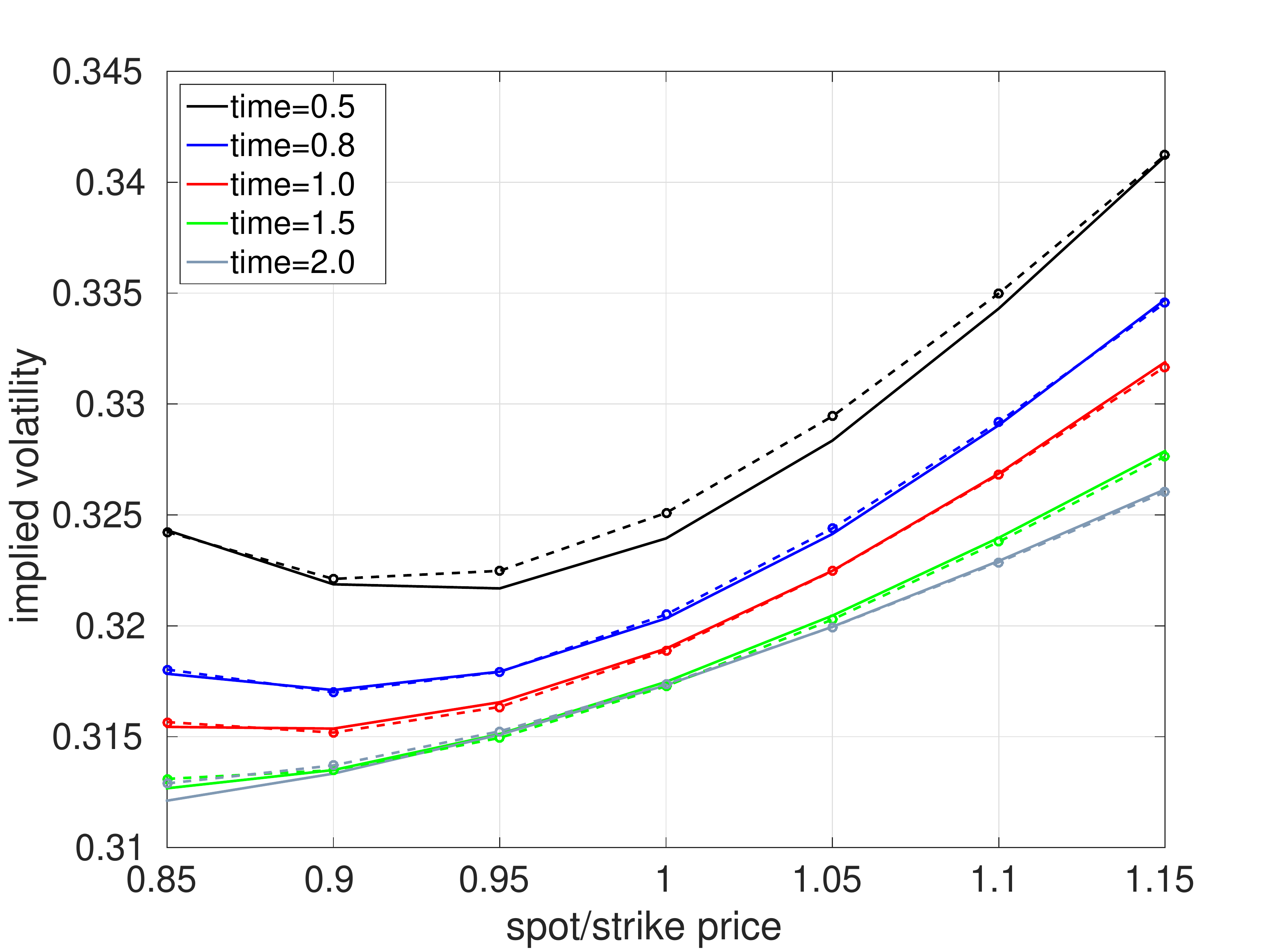}}}%
\subfloat[common jump, $\omega_{ATM}=5000$]{{\includegraphics[width=0.5\textwidth]{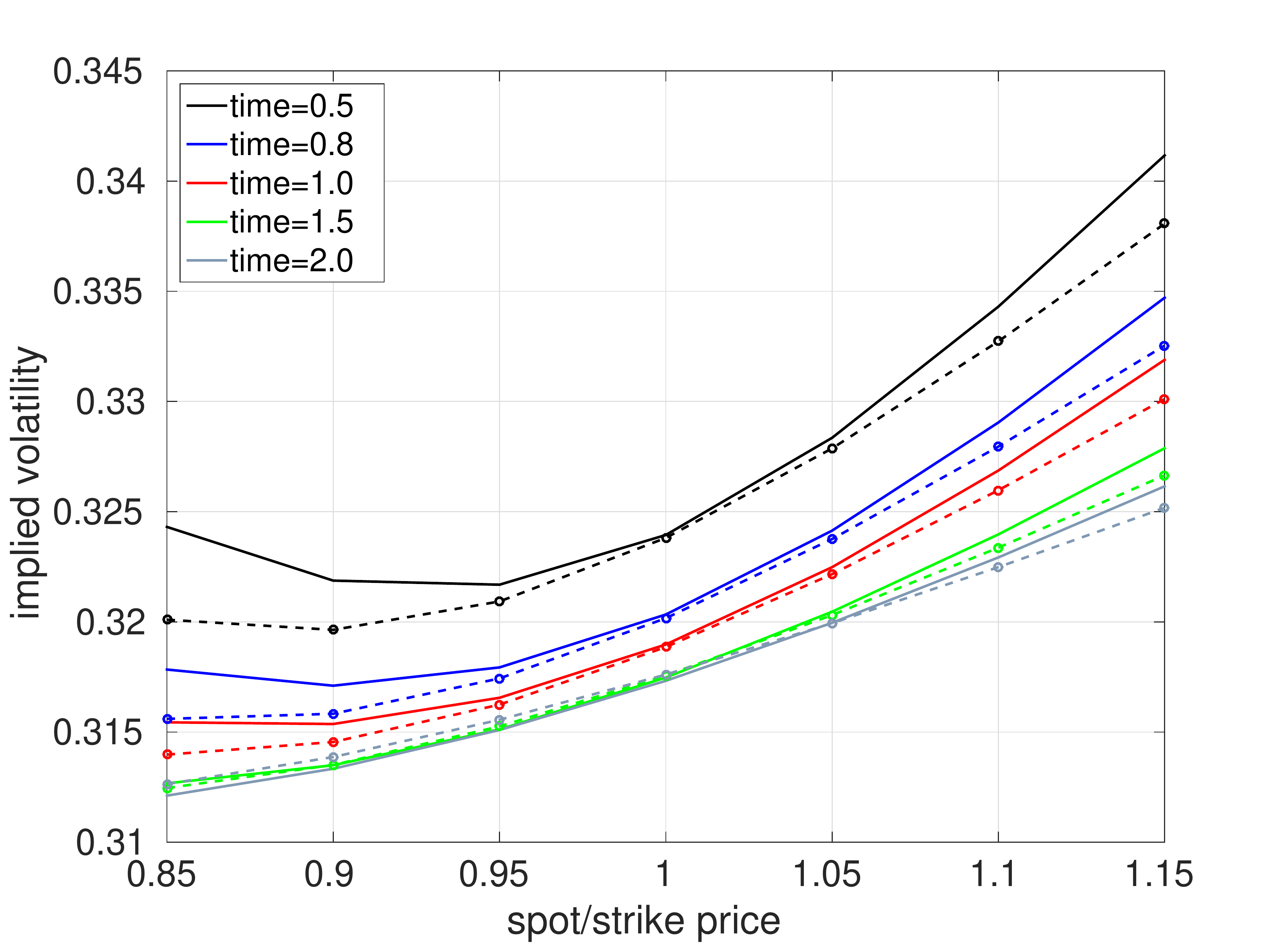}}}%
\caption{Implied volatilities from the 'market' and calibration. The solid lines represent the Bates implied volatilities, while the dashed lines are the calibrated Heston-based volatilities. The impact of weighting ATM options can be seen in the third figure.}%
\label{fig:Bates-Heston IV calibration}%
\end{figure}

There are two sets of experiments, based on either rare jumps or common jumps in the stock price process. Figure \ref{fig:Bates-Heston IV calibration} compares the implied volatility from the Bates model (forward) computations and the CaNN-based Heston implied volatilities. Clearly, when the impact of the jumps is small, the Heston model can accurately mimic the implied volatility generated by the Bates parameters. In this case, many different input parameters for the Bates model will give very similar implied volatility surfaces. With an increasing jump intensity, the deviation between two models can become significant, especially for short maturity options.

In financial practice, a perfect calibration to the ATM options is often required. We can enforce this, by increasing the weights of the ATM options in the objective function. The third figure from Figure~\ref{fig:Bates-Heston IV calibration} and Table \ref{table:Heston-Bates parameter} compare the differences when weighting ATM options in the objective function. The two curves fit very well ATM, however, in this case the total error increases with unequal weighting. The results demonstrate the robustness of the CaNN framework. It is however well-known that the Heston model can not fit short‐maturity market implied volatility very well, and therefore we will also employ a higher-dimensional model, e.g., calibrating directly the Bates model,  which will be discussed in Section \ref{section:bates-cann}.

\subsection{Calibrating the Bates model} \label{section:bates-cann}

In this section, we show the ability of Bates-CaNN to calibrate the Bates model parameters. The Bates model calibration is a higher-dimensional problem, since the Bates model is based on more parameters than the Heston model. The proposed CaNN framework is used to calibrate eight parameters in the Bates model, a setting in which we are dealing with more complex implied volatility surfaces. 

Initially, the Bates-IV-ANN forward pass is trained on the training data set consisting of one million samples that are generated by the Bates model. Compared to the forward pass of the Heston model, merely a different characteristic function is inserted in the COS method, and three additional model parameters have been varied. The Bates-CaNN is employed to calibrate the Bates model, aiming to recover the eight Bates model parameters possibly well. All the samples have equal weight, and the regularization factor is $\bar\lambda = 1.0\times10^{-6}$.

Table \ref{table:Bates parameter weighting} shows an example with high intensity, large variance jumps, for which the Heston model can not capture the corresponding implied volatility accurately. There are still 35 market samples as shown in Table \ref{table:Market parameter range}. This is a challenging problem, estimating eight parameters, and including millions of comparisons between the model  and the market values during calibration.

\begin{figure}[H]
  \centering
  \includegraphics[width=0.8\textwidth]{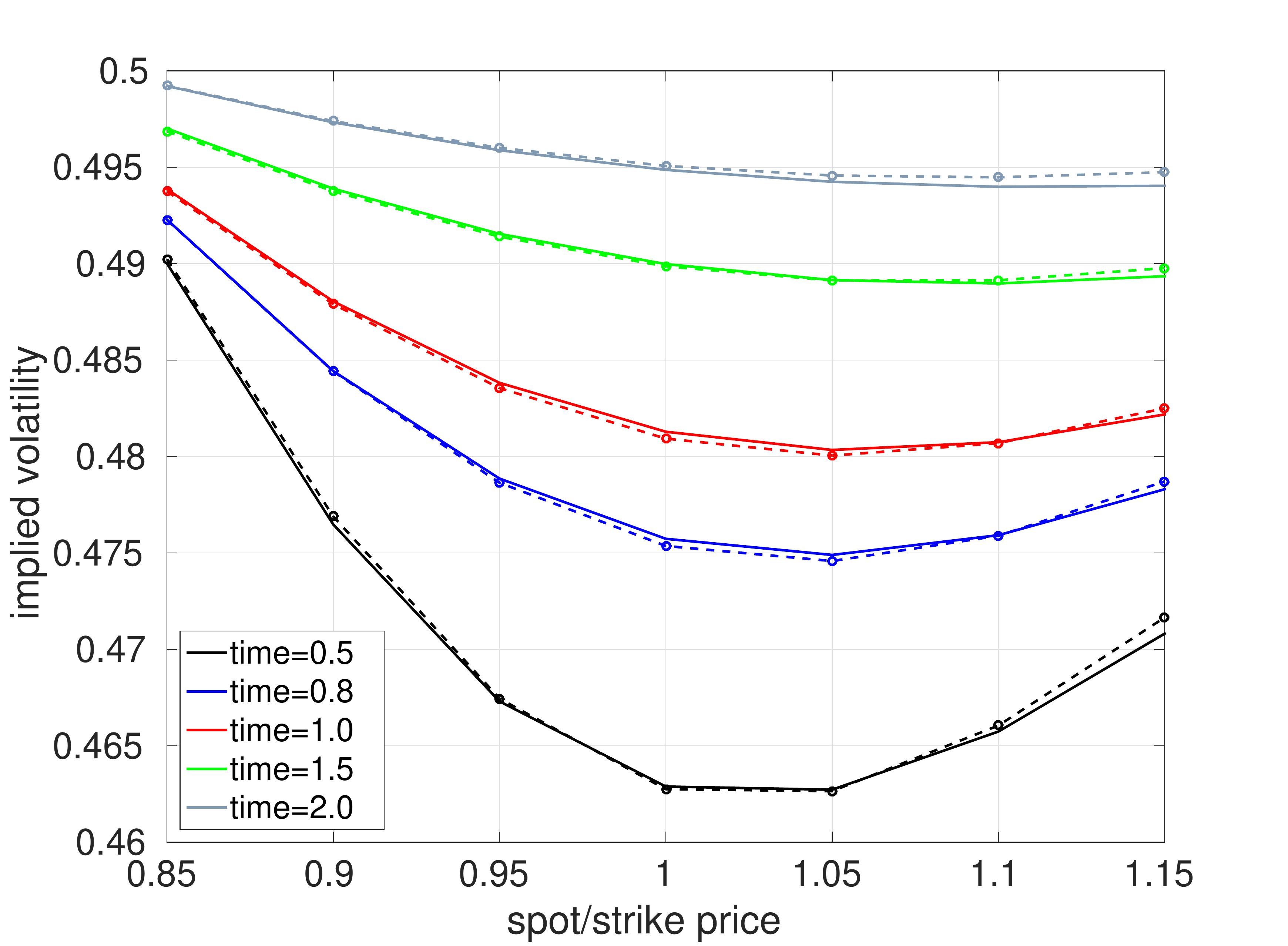}
  \caption{ The solid lines represent the observed implied volatilities, with the dashed lines being the model calibrated ones. This plot shows the result with equal weights and $\bar\lambda=1.0\times10^{-6}$. The random seed is 2 during calibration.}
  \label{fig:Bates intensive jump}
\end{figure}

 Figure \ref{fig:Bates intensive jump} compares the implied volatilities from the synthetic market and the calibrated Bates model. These volatilities resemble each other very well, even when the curvature is high with short time to maturity.

\begin{table}[H]
\caption{The Bates parameters are estimated with Bates-CaNN, by calibrating to a data set (35 samples) generated by the Bates model. In DE, the random seed is 2 and the population size is $10\times N_v=80$.}
\begin{center}
\scalebox{1.0} 
{
 \begin{tabular}{ c|c | c c  }

  \hline
  Parameters         & CaNN Search space  & Bates & Calibrated \\ \hline
  
  Intensity of jumps, $\lambda_J$     & [0, 3.0] & 1.0 & 1.065 \\ 
  
  Mean of jumps, $\mu_J$    & [0, 0.4] & 0.1 & 0.087 \\

  Variance of jumps, $\nu^2_J$    & [0, 0.3]  & $0.4^2$ & 0.146 \\
 \hline
  Correlation, $\rho$     & [-0.9, 0.0]  &-0.3 & -0.228  \\

  Reversion speed, $\kappa$  & [0.1, 3.0]  & 1.0 & 0.598 \\

  Long average variance, $\bar{\nu}$ & [0.01, 0.5]  & 0.1 & 0.128 \\

  Volatility of volatility, $\gamma$ & [0.01, 0.8]  & 0.7 &  0.776 \\  

  Initial variance, $\nu_0$  & [0.01, 0.5]  & 0.1  & 0.102\\ 
  \hline
  Total Squared Error & -& - & 4.95$\times 10^{-6}$
 \\ 
  Function evaluation & -& - & 842800 \\
  Time(seconds)    & - & - & 1.8 \\
  \hline
 \end{tabular}
 }
 \label{table:Bates parameter weighting}
 \end{center}
 \end{table}

\section{Conclusion}

In this work we proposed a machine learning-based framework to calibrate pricing models, in particular focussing on the high-dimensional calibration problems of the Heston and Bates  models. The proposed approach has several favorable features, where an important one is robustness. Without choosing specific initial values, the DE global optimizer prevents the model calibration getting stuck in a local minimum. 

Fast calibration results from several factors. An ANN is efficient in computing the output values for a single input setting. When calibrating, the market data can be computed by ANNs simultaneously. Using DE, during the selection stage, ANNs can calculate a whole population in each generation at once, 
in parallel on a parallel computing architecture.  The numerical experiments show that optimal values can be found within a second even when using a global optimization algorithm. 

Using the ANN-based approach provides new tools to gain insight into the  calibration problem. We used the Hessian matrix to perform a sensitivity analysis, where the sensitivities can efficiently be extracted for large numbers of model parameters. The Hessian matrix also explained why implied volatility, used in our work, is preferred over option prices, used in previous works, from an optimization perspective.

The calibration framework furthermore is generic, and does not require characteristic functions, or explicit gradients of financial models. The number of  market data or to-calibrate parameters is also flexible.  With this framework, the model can be extended to multiple quantities, e.g., calibrating to both option prices and implied volatility. To conclude, the ANN combined with DE provides an efficient and accurate framework for calibrating financial models.

To look forward, the above ANN calibration process does not rely on the quality of the initial guess. However, because the market does not change dramatically in a short time period, it may make sense to take the last available values as starting point of the calibration in future work. There are several possible strategies for the calibration framework in this situation. One is switching to the gradient-based local optimization algorithms and another one is narrowing the search space of the DE, which will further reduce the computational time considerably.  Further future improvements include combining gradient-based optimization with the DE, since the gradient information is readily accessible. It is also feasible to employ a small neural network to reduce the computing time, like in the paper \citep{deeplearningvol:2019} which builds a three-hidden-layers ANNs and each layer has 30 nodes during the calibration. 

\section{Acknowledgements}
The authors would like to acknowledge  the many fruitful discussions with Professor Sander M.Bohte, and the financial support from the China Scholarship Council (CSC). This research was conducted using the supercomputer Little Green Machine II in the Netherlands.

\bibliographystyle{apalike} 
\bibliography{references}

\appendix
\section{COS Pricing Method}\label{sec:cos}
 
Based on the Feynman-Kac Theorem, the solution of the governing option valuation PDEs is given by the risk-neutral valuation formula,
\begin{equation*}
V(t_0,x,\nu)=e^{-r\Delta t}\int_{-\infty}^{\infty}V(T,y,\nu)f(y\vert x)dy,
\end{equation*}
where $V(t,x,\nu)$ is the option value, and $x, y$ are increasing functions of the underlying at $t_0$ and $T$, respectively. To get to the COS formula, we truncate the integration range, so that
\begin{equation}
V(t_0,x,\nu)\approx e^{-r\Delta t}\int_{a}^{b}V(T,y,\nu)f(y\vert x)dy.
\label{a1}
\end{equation}
with $\vert\int_{\mathbb R} f(y\vert x)dy-\int_{a}^{b}f(y\vert x)dy\vert<TOL$.
 
The density function of the underlying is then approximated by means of the characteristic function with a truncated Fourier cosine expansion, as follows:
\begin{equation}\label{den}
f(y \vert x) \approx \frac{2}{b-a}\sump_{k=0}^{N-1} Re(\hat f(\frac{k\pi}{b-a};x)\exp{(-i\frac{ak\pi}{b-a})})\cos{(k\pi\frac{y-a}{b-a})},
\end{equation}
where $Re$ means taking the real part of the expression in brackets, and $\hat f(\omega;x)$ is the characteristic function of $f(y\vert x)$ defined as below
\begin{equation}\label{cf}
\hat f(\omega;x)=\mathbb{E}(e^{i\omega y}\vert x).
\end{equation}
The prime at the sum symbol in~(\ref{den}) indicates that the first term in the expansion is multiplied by one-half.
Replacing $f(y\vert x)$ by its approximation~(\ref{den}) in~(\ref{a1}) and interchanging the order of integration and summation, gives us the COS algorithm to approximate the value of a European option, as below:
\begin{equation}\label{cos}
V(t_0,x,\nu)=e^{-r\Delta t}\sump_{k=0}^{N-1}Re(\hat f(\frac{k\pi}{b-a};x)e^{-ik\pi\frac{a}{b-a}})H_k,
\end{equation}
where
\begin{equation}\label{vk}
H_k=\frac{2}{b-a}\int_a^bV(T,y,\nu)\cos{(k\pi\frac{y-a}{b-a})}dy
\end{equation}
is the Fourier cosine coefficient of $H(t,y)=V(T,y,\nu)$, which is available in closed-form for several European option payoff functions.
 
Formula~(\ref{cos}) can be directly applied to calculate the value of European option, it also forms the basis for the pricing of Bermudan options.
 
The COS algorithm exhibits an exponential convergence rate for all processes whose conditional density $f(y\vert x) \in C^{\infty}((a,b) \subset \mathbb R)$.
The size of the integration interval $[a, b]$ can be determined with help of the cumulants.
\end{document}